\documentclass[aps,11pt,groupedaddress,nofootinbib]{revtex4-1}
\pdfoutput=1

\usepackage{mathtools}
\usepackage{graphicx}
\usepackage{float}
\usepackage[T1]{fontenc} % if needed
\usepackage{slashed}
\usepackage{amsmath}
\usepackage{color}
\usepackage{amsfonts}
\begin{document}
\allowdisplaybreaks

\title{Spinor driven cosmic bounces and their cosmological perturbations}

\author{Shane Farnsworth}
\author{Jean-Luc Lehners}

%\thanks{}
\affiliation{Max Planck Institute for Gravitational Physics (Albert Einstein Institute), 14476 Potsdam, Germany}

\author{Taotao Qiu}
\affiliation{College of Physical Science and Technology, Central China Normal University, Wuhan, People's Republic of China. \\}


%\date{\today}

\begin{abstract}
\vspace{.5cm}
When coupling fermions to gravity, torsion is naturally induced. We consider the possibility that fermion bilinears can act as a source for torsion, altering the dynamics of the early universe such that the big bang gets replaced with a classical non-singular bounce. We extend previous studies in several ways: we allow more general fermion couplings, consider both commuting and anti-commuting spinors, and demonstrate that with an appropriate choice of potential one can easily obtain essentially arbitrary equations of state, including violations of the null energy condition, as required for a bounce. As an example, we construct a model of ekpyrotic contraction followed by a non-singular bounce into an expanding phase.  We analyze cosmological fluctuations in these models, and show that the perturbations can be rewritten in real fluid form. We find indications that spinor bounces are stable, and exhibit several solutions for the perturbations. Interestingly, spinor models do not admit a scalar-vector-tensor decomposition, and consequently  some types of scalar fluctuations can act as a source for gravitational waves already at linear order. We also find that the first order dynamics are directionally dependent, an effect which might lead to distinguished observational signatures.
\end{abstract}

%\keywords{}

\maketitle

\tableofcontents
\section{Introduction}
\label{Sec_Introduction}

Any successful model of cosmology is required to explain the large scale properties of our universe, including its near homogeneity, isotropy and flatness, as well as the almost
scale-invariant spectrum of its primordial density perturbations. Our current understanding is that of an expanding universe initiated at a big bang singularity, at which point our usual effective description given in terms of general relativity breaks down. It is natural to wonder therefore how this initial singularity might be resolved, especially in light of recent results showing that a replacement of the big bang by regular semi-classical geometries \cite{Vilenkin:1983xq,Hartle:1983ai} does not work \cite{Feldbrugge:2017kzv,Feldbrugge:2017fcc,DiazDorronsoro:2017hti,Feldbrugge:2017mbc}. An attractive possibility, and the basis for this paper, is to replace the big bang with a bounce in which the current expanding phase of our universe emerges from a prior period of contraction. 

While a cosmological bounce may be induced by quantum
gravity effects when the scale factor of the universe shrinks to near the Planck scale \cite{Gielen:2015uaa,Chen:2016ask,Bramberger:2017cgf}, in this paper we are interested in classical non-singular bouncing scenarios. In such scenarios the contraction of the universe stops and reverses into an expanding phase at a finite value of the scale factor `$a$' when a classical description remains valid. In this way it should be possible to follow the entire cosmological  evolution through the bounce using the well understood  framework of general relativity and effective field theory \cite{Cai:2016thi, Creminelli:2016zwa, Cai:2017tku}.

According to the singularity theorems
of Penrose and Hawking~\cite{HP70}, under rather broad assumptions the null energy condition (NEC) must be violated in order to obtain a non-singular bounce. This usually requires the introduction of some sort of NEC violating exotic matter, such as a scalar field that undergoes ghost condensation (see e.g. \cite{Creminelli:2006xe,Creminelli:2007aq,BKO07,Lehners:2011kr}) or models involving Galileon fields (see e.g. \cite{Qiu:2011cy,Easson:2011zy,Cai:2012va,Qiu:2013eoa,Ijjas:2016tpn,Ijjas:2016vtq, Qiu:2015nha, Wan:2015hya}; such models can also be embedded into supergravity \cite{Koehn:2012te,Koehn:2013upa}). While a scalar condensate phase is not difficult to achieve on its own, the situation becomes much more restrictive once observational and stability requirements are taken into consideration~\cite{Battarra:2014tga,Koehn:2015vvy,deRham:2017aoj}. The purpose of this paper is to see if a more desirable outcome might be achieved by making use of fermionic rather than scalar matter. Such an approach is sensible to consider for two reasons: the first being of course the natural predominance of fermionic matter in the standard model of particle physics (as well as the comparable dearth of fundamental scalar fields). The second, as we will discuss briefly in the bulk of this work, is the relative ease with which any desirable equation of state is achievable using spinor fields~\cite{ArmendarizPicon:2003qk}.

In this paper we explore models of gravity with torsion coupled to spinors \cite{Hehl:1976kj}. Of particular interest are models in which the  torsion is non-dynamical and sourced by the spinor content. Such models arise naturally when viewing general relativity as a gauge theory, more specifically as having the gauge symmetries of reparameterisations and local Lorentz transformations (see e.g. \cite{Ortin:2015hya}) -- it is then also known as the Cartan-Sciama-Kibble theory \cite{Kibble:1961ba,Sciama:1964wt,Stelle:1979va}. In order to be able to treat the spinor bilinear sources classically, one must make some assumptions about the nature of the spinors -- here we will consider both commuting and anti-commuting spinors, and will discuss the respective assumptions in some detail. In this framework the usual dynamics of Einstein gravity is recovered under most circumstances of interest, and the effects of torsion only become relevant in regions of extreme spinor density. The idea is that in a contracting universe, as the scale factor  drops the spinor density increases until eventually a bounce is precipitated \cite{Poplawski:2012ab,Poplawski:2011jz,Magueijo:2012ug,Alexander:2014eva,Alexander:2014uaa}. We will show how one is readily able to construct backgrounds which not only undergo a bounce, but which also accommodate other interesting dynamics outside the bouncing phase, such as inflation or ekpyrosis. 

As usual, a study of cosmological perturbations is crucial both in order to assess the viability of these models, and to see if one might be able to distinguish them using cosmological observations. We investigate linear perturbation theory in some detail, showing for instance that the linearized equations of motion may be cast in real fluid form. Our main results include a derivation of the equations of motion for models that include more general spinor-torsion couplings than appear elsewhere in the literature, the realization of the absence of a scalar-vector-tensor decomposition, the derivation of several solutions for the perturbations, and the identification of directionally dependent perturbations. These features are in direct contrast with the known results regarding perturbations for non-singular bounces sourced by scalar field matter~\cite{Battarra:2014tga}, and indicate that spinor bounces may indeed have their own specific observational signatures. 

We organize the paper as follows: we first introduce our model and present the equations of motion in section \ref{Sec_Model}. Then, in section \ref{Sec_Background} we discuss how the restriction to simple cosmological metrics considerably simplifies the dynamical equations and allows for bouncing solutions. The core of the paper is in section \ref{Sec_Perturbations}, in which we analyze the cosmological perturbations of these models by studying the linearised equations of motion. We discuss our results in section \ref{Sec_Discussion}. In the extensive appendix we present our conventions and provide details regarding both the derivation of the equations of motion and the construction of perturbation theory.

\section{The Model}
\label{Sec_Model}

%\subsection{The action}
%\label{Sec_Model_action}
In our model the action $S$ will be split into two parts: the gravitational sector $S_G$, and the matter sector $S_\Psi$. Our goal is to  explore the effect of torsion on gravitational dynamics, and so we begin by first introducing the most general gravitational action. Anticipating the introduction of fermionic matter we work within the first order  formalism, written in terms of the frame field $e^I = e_\mu^I dx^\mu$, and the Lorentz connection $\omega^{IJ} = \omega^{IJ}_{~~\mu}dx^\mu$. Following the effective field theory approach~\cite{Freidel:2005sn}, attention is restricted  to Lagrangians which are generally covariant, locally Lorentz invariant, and polynomial in the basic fields and their derivatives. Under such restrictions there are only six possible terms that can be written down to leading order, three of which are topological and will not be considered~\cite{Rezende:2009sv}. Of the three bulk terms, one is given by the cosmological constant, and will be included in the matter action $S_\Psi$. The remaining two possible terms we take for our  gravitational action:
\begin{align}
S_{G} &= \kappa\int  (\epsilon_{IJKL} +\tfrac{2}{\gamma}
\eta_{I[K}\eta_{J]L})e^Ie^J R^{KL},\label{Eq_Act_Grav}
\end{align}
where $\kappa = 1/32\pi G$. Equation~\eqref{Eq_Act_Grav} is known as the Cartan-Holst action, while the coupling constant $\gamma$ is known as the Immirzi parameter. Note that in a theory without torsion the second term in Eq.~\eqref{Eq_Act_Grav} is identically zero due to the symmetries of the Riemann tensor, in which case Eq.~\eqref{Eq_Act_Grav} reduces to the familiar Einstein-Hilbert action.

We next need to introduce a source for the torsion in our model. We consider  Dirac spinors, which we couple into our model by including the following matter action:
\begin{align}
\begin{split}
S_\Psi &= \tfrac{i}{2.3!}\int \epsilon_{IJKL}e^Ie^Je^K(\overline{\Psi}\gamma^L D\Psi - \overline{D\Psi}\gamma^L\Psi) -\tfrac{1}{4!} 
\int \epsilon_{IJKL}e^Ie^Je^Ke^L U(\overline{\Psi}\Psi)\\
&+\tfrac{1}{4}\int\epsilon_{IJKL}e^Je^K(De^I)\Theta^L +\tfrac{1}{4}\int\eta_{I[K}\eta_{J]L}e^Je^K(De^I)\Omega^L,\end{split}\label{Eq_Act_Ferm}
\end{align}
where $D$ is the covariant exterior derivative with torsion, and $\Psi$ is a Dirac spinor. The potential $U$ is an arbitrary function of the spinor bilinear $\overline{\Psi}\Psi$, and might include for example a cosmological constant tern. For compactness of notation we have defined the spinor currents $\Theta^L \equiv (\alpha V^L + \beta A^L)$, and $\Omega^L \equiv (\tau V^L + \lambda A^L)$, where $\alpha,\tau,\beta,$ and $\lambda$ are arbitrary coupling constants, and the vector and axial spinor bilinears are given respectively by:
\begin{align}
V^L &= \overline{\Psi}\gamma^L\Psi, & A^L &= \overline{\Psi}\gamma_5\gamma^L\Psi.
\end{align}
Our full action is given by $S = S_G + S_\Psi$. Note that we have implemented general couplings between torsion and the vector and axial currents in order to encapsulate the various models present elsewhere in the literature. By setting $\tau=\lambda=0$ we recover the matter action introduced in~\cite{Magueijo:2012ug}, while setting $\beta =\tau=\lambda=0$ we recover the matter action given for example in~\cite{Freidel:2005sn,Randono:2005up,Khriplovich:2005jh}. Turning off all torsion couplings  $\alpha =\beta =\tau=\lambda=0$ recovers the matter action discussed for example in~\cite{Ellis:2011mz}.

%\subsection{Equations of motion}
%\label{Sec_Model_EOM}

Having constructed  the action provided in Eqs.~\eqref{Eq_Act_Grav} and~\eqref{Eq_Act_Ferm}, we can now determine the corresponding equations of motion.  In order to do so we take the vierbein $e$, spin connection $\omega,$ and spinor $\Psi,$ as our fundamental fields, and vary the action with respect to each in turn. As the calculation itself is rather  long and involved, we will simply  outline our final results. In the appendix we present the trickier parts of the calculation, and also provide the full calculation for the simpler case in which the Holst term in the action is removed and the torsion couplings are all switched off, i.e.  $\alpha = \beta = \tau = \lambda=0$.
 
We begin by considering the variation of the action $S = S_G+S_\Psi$ with respect to the spin connection $\omega^{MN}$, which yields the following equation of motion:
\begin{align}
\begin{split}
2\kappa(\epsilon_{IJMN}+\tfrac{2}{\gamma}\eta_{I[M}\eta_{N]J})(De^I)e^J &= -\tfrac{1}{4!}\epsilon_{IJKL}e^Ie^Je^K\varepsilon^{DL}_{\bullet\bullet MN}A_D - \tfrac{1}{4}\epsilon_{[M|JKL}e^Je^Ke_{|N]}\Theta^L\\
&~~~-\tfrac{1}{4}e^Je^K e_{[N}\eta_{M]K}\eta_{JL}\Omega^L.
\end{split}
\label{FullMod_EOM_omega_1}
\end{align}
In its current form Eq.~\eqref{FullMod_EOM_omega_1}  is rather opaque to interpretation. We can however make progress  by solving it to obtain an algebraic expression for the contortion $C^{TXS},$ defined via $C^I_{MN}e^Ne^M=De^I.$ Full details are provided in the appendix, and the resulting expression for the contortion is
\begin{align}
C^{TXS}&= \frac{\gamma^2}{8\kappa(1+\gamma^2)}\Big[\tfrac{1}{2}\varepsilon^{QXST}(\tfrac{1}{\gamma}\Theta_Q-(A_Q+\Omega_Q))
+\eta^{S[T}\delta^{X]}_A(\Theta^A+\tfrac{1}{\gamma}(A^D+\Omega^D))\Big].
\label{FullMod_EOM_omega_2}
\end{align}
Our result generalises the work found for example in~\cite{Magueijo:2012ug,Freidel:2005sn,Randono:2005up,
Khriplovich:2005jh,Ellis:2011mz}. The reader should take care when comparing between papers however, as there are a range of different sign conventions being used. Notice that our expression for the contortion is algebraic, and depends only on the vector and axial spinor densities. In particular, if we had not coupled spinors into our model the contortion would have been identically zero. 

We next vary the action with respect to the spinor $\Psi$ to obtain the following curved space Dirac equation:
\begin{align}
\begin{split}
\tfrac{i}{3!}\epsilon_{IJKL}\varepsilon^{IJKM}\gamma^L\widetilde{D}_M\Psi
&=
-\tfrac{1}{4}\epsilon_{IJKL}\varepsilon^{PQJK}C^I_{\bullet QP}\frac{\delta \Theta^L}{\delta \overline{\Psi}} -\tfrac{1}{4}\eta_{I[K}\eta_{J]l}\varepsilon^{PQJK}C^I_{\bullet QP}\frac{\delta \Omega^L}{\delta \overline{\Psi}}\\
&~~~+\tfrac{i}{8.3!}\epsilon_{IJKL}\varepsilon^{IJKM}C_{ABM}\gamma^L[\gamma^A,\gamma^B]\Psi + 
\frac{\delta U}{\delta \overline{\Psi}},
\end{split}\label{FullMod_EOM_omega_3} 
\end{align}
where we have used tildes to indicate when a term is taken to be torsion free (See Appendix~\ref{convens}). The Dirac equation can be simplified considerably by making use of the expression for the contortion derived in Eq.~\eqref{FullMod_EOM_omega_2}. A long calculation leads to
\begin{align}
ie^\mu_L\gamma^L\widetilde{D}_\mu\Psi &=\frac{\delta W}{\delta\overline{\Psi}}\label{FullMod_EOM_omega_4},
\end{align}
where the effective potential $W$ is defined by
\begin{align}
W &= U(E) + \xi_{AA}A^IA_I + 2\xi_{VA}V^IA_I + \xi_{VV}V^IV_I,
\label{FullMod_EOM_omega_5}
\end{align}
 and where
\begin{align}
\begin{split}
\xi_{AA} &= - \frac{3\pi G \gamma^2}{2(1+\gamma^2)}(\tfrac{2}{\gamma}\beta(1+\lambda)+\beta^2 - (1+\lambda)^2),\\
\xi_{AA} &= - \frac{3\pi G \gamma^2}{2(1+\gamma^2)}(\tfrac{2}{\gamma}\alpha\tau + \alpha^2-\tau^2),\\
\xi_{AV} &=- \frac{3\pi G \gamma^2}{2(1+\gamma^2)}(\alpha(\beta + \tfrac{1}{\gamma}(1+\lambda)) + \tau(\tfrac{1}{\gamma}\beta - (1+\lambda)).
\end{split}
\end{align}
Finally,  the Einstein equations are obtained by varying the action with respect to the vierbein,
\begin{align}
\begin{split}
0&= 2\kappa(\epsilon_{SJKL} + \tfrac{2}{\gamma}\eta_{SK}\eta_{JL})e^J(\widetilde{R}^{KL} + \widetilde{D}C^{KL} + C^K_{~P}C^{PL})\\
&~~~+\tfrac{i}{4}\epsilon_{SJKL}e^Je^KX^L- \tfrac{1}{3!}\epsilon_{SJKL}e^Je^ke^LU + \tfrac{1}{4}\epsilon_{SJKL}e^Je^K D \Theta^L\\
&~~~+\tfrac{1}{2}(\eta_{S[K}\eta_{J]L}+ \eta_{J[K}\eta_{S]L})(De^J)\Omega^L + \tfrac{1}{4}\eta_{SK}\eta_{JL}e^Je^KD\Omega^L,
\end{split} \label{FullMod_EOM_omega_6}
\end{align}
where $X^L = (\overline{\Psi}\gamma^L
D\Psi - \overline{D\Psi}\gamma^L\Psi)$ and where we are once again using the tildes to indicate when a quantity is torsion free. The Einstein equations appear unfamiliar in this first order form, but can be re-expressed in second order form. After another lengthy but rather straightforward calculation requiring Eq.~\eqref{FullMod_EOM_omega_2} and repeated use of the identities given in Eq.~\eqref{Gamma_identities},  the following compact form may be obtained:
\begin{align}
\begin{split}
4\kappa \widetilde{G}_{\mu\nu} &= -\tfrac{i}{2}\left[e_{a(\mu}\widetilde{X}_{\mu)}^a-\widetilde{X}g_{\mu\nu}
\right]- g_{\mu\nu} W\\
&~~~+\tfrac{1}{8}e_{a\nu}e_{b\mu}\left[\overline{\Psi}[\gamma^a,\gamma^b]\gamma^c\widetilde{D}_c\Psi -\overline{ \widetilde{D_c}\Psi}\gamma^c[\gamma^a,\gamma^b]\Psi\right],
\end{split}\label{FullMod_EOM_omega_7}
\end{align}
where we have defined $\widetilde{X}^L_\tau = (\overline{\Psi}\gamma^L
\widetilde{D}_\tau\Psi - \overline{\widetilde{D}_\tau\Psi}\gamma^L\Psi)$. Notice that the last term on the RHS is not symmetric in its indices, which appears to be in conflict with the symmetries of the torsion free Einstein tensor. However, by making use of the Dirac equation this term is found to be identically zero on shell.

\section{Background Cosmology}
\label{Sec_Background}

In this section we find bouncing, cosmological background solutions for the equations of motion which were derived above. In order to do so we make two simplifying assumptions: (i) we impose a `classicality' assumption on spinor bi-linears, in which we view spinor pairs $\overline{\Psi}\Psi$ as forming a classical bosonic condensate, and (ii) we take a flat Friedman-Lema\^{i}tre-Robertson-Walker (FLRW) ansatz for the background metric. With these two assumptions the equations of motion simplify rather dramatically and even allow for analytic solutions as we will describe.

\subsection{Classicality conditions}
\label{Sec_Background_Classical}

Having derived the equations of motion in Sec.~\ref{Sec_Model}, our next goal is to interpret them.  The usual approach in the literature has been to view  the equations of motion as operator equations, and assume that the classical gravitational field that we observe is sourced by the expectation value of spinor bilinears such as $\langle A^I\rangle$ and $\langle V^I\rangle$. Taking this approach leads to an ambiguity however when considering the four point spinor interaction terms present in Eq.~\eqref{FullMod_EOM_omega_4}. In particular, starting with the first order formalism and solving for the classical contortion, one obtains interaction terms of the form $\langle A^I\rangle \langle A_I\rangle$. On the other hand, if one had instead started from the second-order formalism with quartic interactions, then contributions of the form $\langle A^I A_I\rangle$ would  be obtained. The problem is that  $\langle A^I\rangle \langle A_I\rangle$ and $\langle A^I A_I\rangle$ are not in general  equal~\cite{Dolan:2009ni,Magueijo:2012ug}. 

To avoid any ambiguity, previous authors have restricted their attention to so called `classical spinors' $\Psi_{cl} = \langle \Psi \rangle$~\cite{ArmendarizPicon:2003qk,Magueijo:2012ug}. Classical spinors are defined as the expectation value of the operator $\Psi$ in a state such that $f(\langle \Psi\rangle) \simeq \langle f(\Psi)\rangle$ for any function $f$. In practice the classical spinor assumption is extremely stringent, and amounts to  describing $\Psi_{cl}$  as a four component object with complex entries. While it is fine to presume the existence of classical spinors as an effective description of nature, this assumption is not well motivated by known physics. The fermions of the standard model of particle physics are quantum fields, and  due to the Pauli exclusion principle it is not well understood when they might be treated consistently as a classical spinor condensate. Thankfully it is really not necessary to impose any classicality conditions directly on the spinors in our model. It is only spinor bilinears which appear in the Einstein equations, and similarly the Dirac equation may be re-expressed in projected form in terms of bilinears. A weaker  classicality assumption which one might then consider, is to ask instead that the variance of the various spinor bilinears is small, i.e.
\begin{align}
\langle A^IA_I\rangle \simeq \langle A^I\rangle\langle A_I\rangle\label{Classicaility_Cond},
\end{align}
together with similar relations for the other bilinear terms present in the model.

Once this `variance' assumption has been made  the Fierz identity can be used to further simplify the form of the potential given in~\eqref{FullMod_EOM_omega_5}.  In four dimensions the generalized Fierz identity is given by~\cite{Ortin:2015hya}:
\begin{align}
\begin{split}
s
(\overline{\lambda}M\chi)(\overline{\psi}N\phi) &=
-\tfrac{1}{4}(\overline{\lambda}MN\phi)(\overline{\psi}\chi)+\tfrac{1}{4}(\overline{\lambda}M\gamma^aN\phi)(\overline{\psi}\gamma_a\chi)-\tfrac{1}{4}(\overline{\lambda}M\gamma_5N\chi)(\overline{\psi}\gamma_5\phi)\\
&~~~+\tfrac{1}{8}(\overline{\lambda}M\gamma^{[ab]}N\phi)(\overline{\psi}\gamma_{[ab]}\chi)-\tfrac{1}{4}(\overline{\lambda}M\gamma_5\gamma^a N\phi)(\overline{\psi}\gamma_5\gamma_a\chi)
\end{split}\label{Id_Fierz}
\end{align}
where $\lambda, \chi, \psi,$ and $\phi$ are Dirac Spinors, and where the sign $s$ depends on the spin statistics chosen. For commuting spinors, $s = -1$ and we can immediately derive the following identities:
\begin{align}
\begin{split}
\langle A_I A^I\rangle = -\langle V_I V^I\rangle &= E^2 + B^2,\\
\langle A_I V^I\rangle &=0,
\end{split}\label{Eq_Bilinearex_1}
\end{align}
where we have defined the following densities $E = \langle\overline{\Psi}\Psi\rangle$ and $iB = \langle\overline{\Psi}\gamma_5\Psi\rangle$. For commuting spinors the effective potential given in Eq.~\eqref{FullMod_EOM_omega_5} therefore simplifies considerably:
\begin{align}
W &= U(E)+ \xi(E^2+B^2),
\end{align}
where
\begin{align}
\xi = - \frac{3\pi G \gamma^2}{2(1+\gamma^2)}\big[(\tfrac{2}{\gamma}\beta(1+\lambda)+\beta^2 - (1+\lambda)^2)-(\tfrac{2}{\gamma}\alpha\tau + \alpha^2-\tau^2)\big].
\end{align}
This is the same form for the potential found by those authors who impose the  so called `classical spinor' assumption~\cite{ArmendarizPicon:2003qk,Magueijo:2012ug}. In practice, since we will ultimately only ever be dealing with spinor bilinears, what we really mean by `commuting spinors' is that the following two conditions hold: (i) first we ask that all quartic spinor terms may be expressed as the square of bilinear terms, i.e. of the form $\langle V^I\rangle \langle V_I\rangle$. This requires either the primacy of the first order formalism, or that the variance condition from Eq.~\eqref{Classicaility_Cond} hold. And (ii) that the conditions derived in Eqs.~\eqref{Eq_Bilinearex_1} for commutative spinors, hold. As we will show, under these two assumptions the equations of motion simplify dramatically, allowing us to find a number of very interesting analytic solutions even at linear order in perturbations.

It is of course physically more interesting to consider the case of anti-commuting spinors for which $s=1$. In this case the Fierz identity together with our classicality assumption on bilinears yields the following relation: 
\begin{align}
\tfrac{1}{2}(\langle V^IV_I\rangle -\langle A^IA_I\rangle) = E^2 + B^2,\label{Eq_Bilinearex_3}
\end{align}
where we note the sign change in front of square of the axial and vector currents.  Since we will only ever deal directly with spinor bilinears, what we mean by `non-commuting' spinors  is that we will be using~\eqref{Eq_Bilinearex_3} in place of the conditions given in Eqs.~\eqref{Eq_Bilinearex_1} for commuting spinors. For anti-commuting spinors our analysis does not require us to impose any restriction at all on the variance of bilinears. It is enough to presume the first order formalism as fundamental, in which case all quartic spinor terms in the classical action are considered to be of the form $\langle A_I\rangle \langle A^I\rangle$.

\subsection{Flat FLRW and commuting spinors} \label{Sec_Background_FLRW}

Our next goal is to construct background solutions which undergo a cosmological bounce. As we demonstrate, bouncing solutions may be readily obtained both for commuting, and anti-commuting spinors. We begin in this section with commuting spinors, which are simpler to deal with computationally. Our procedure is to assume the line element of FLRW spacetime, and then given this assumption check for consistency of the Dirac and Einstein equations given in Eqs.~\eqref{FullMod_EOM_omega_4} and \eqref{FullMod_EOM_omega_7}.  Notice however that it is by no means a foregone conclusion that it will be possible to find cosmological solutions. As can be seen from Eq.~\eqref{Eq_Bilinearex_1}, under our classicality assumption the axial current $A_I$ is spacelike. It appears therefore that spinor fields pick out a preferred direction in spacetime, violating Lorentz invariance and potentially conflicting  with the isotropy assumption of the background metric. As was shown by Isham and Nelson in~\cite{Isham:1974ci} this fear is indeed realised in most, but not all, cases. In most cases, once an FLRW background solution is selected for the metric, the equations of motion force the axial spinor current, and therefore the spinor itself, to be identically zero. For flat FLRW however, there is no such obstruction  and consistent solutions can be found in which the metric remains isotropic despite $A_I$ being anisotropic.

 In this paper we work with the flat FLRW line element expressed in physical time as
\begin{align}
ds^2 = -dt^2  + a(t)^2\delta_{ij}dx^idx^j\,,
\end{align}
and we will denote the Hubble rate by $H = \dot{a}/a$.  Given this choice of line element, it is natural to take the classical spinor $\Psi$ to have no spatial dependence. In this case, a short calculation shows the Einstein and Dirac equations to be:
\begin{subequations}
\begin{align}
12\kappa H^2 &= \tfrac{i}{2} (\overline{\Psi}\gamma^0\dot{\Psi}-
\dot{\overline{\Psi}}\gamma^0{\Psi}) -[\xi (E^2+B^2) + U'E-U]\nonumber\\
&= \left[U+\xi( E^2+B^2)\right],
\label{Eq_Back_Fried1}\\
-4\kappa(2\dot{H} + 3H^2)&= 
[\xi (E^2+B^2) + U'E - U],\label{Eq_Back_Fried2}\\
%
%  %
%
%
%
%
\gamma^0  \partial_0 \Psi  + \tfrac{3}{2} \gamma^0 H\Psi&= 
-i[(U'+2\xi E)\Psi -2i\xi B \gamma_5\Psi],\label{Eq_Back_Dir1}
\end{align}
\end{subequations}
where we have assumed that the potential $U$ is a function of $E$ only, and the `prime' refers to differentiation with respect to $E$. We have made use of the Dirac equation in order to obtain the second line of~\eqref{Eq_Back_Fried1}. 

Notice that because of homogeneity and isotropy of the background, the stress energy tensor on the RHS of the Einstein equations is necessarily of the perfect fluid form
\begin{align}
T^\mu_{\nu} = u^\mu u_\nu(P+\rho) + \delta^\mu_{\nu}P,
\end{align}
where because of homogeneity the pressure and density are functions of time only, i.e. $\rho = \rho(\tau)$ and $P = P(\tau)$, and because of isotropy the fluid is at rest in the background universe: $u_{\mu} = \{-1,0,0,0\}$.  From Eqs.~\eqref{Eq_Back_Fried1} and~\eqref{Eq_Back_Fried2}  we read off immediately that
\begin{align}
P &=\xi (E^2+B^2) + U'E - U, & \rho &=  \xi (E^2+B^2)  +U\,,
\end{align}
while we can use the Dirac equation~\eqref{Eq_Back_Dir1} to obtain the usual conservation equation
\begin{align}
\dot{\rho} = -3H(P+\rho).
\end{align}
Notice that  the spinor field can accommodate any desired behavior for its energy density and equation of state by a judicious choice of potential $U$~\cite{ArmendarizPicon:2003qk}.

As might be expected the RHS of the Einstein equations is expressed entirely in terms of spinor bilinears. It will be useful therefore to also re-express the Dirac equation in projected form, written entirely in terms of spinor bilinears.  In order to do this, first notice that in addition to the bilinears $E$ and $B$ defined below Eq.~\eqref{Eq_Bilinearex_1},  there are six other possible  (non-independent) hermitian spinor bilinears which can be constructed from a single background spinor $\Psi$. The full list is given by:
\begin{align}
\begin{split}
iB &= \overline{\Psi}\gamma_5\Psi,\\
iC^i &= \overline{\Psi}\gamma^0\gamma^i\Psi,
\end{split}
&
\begin{split}
E &=\overline{\Psi}\Psi,\\
V^i &=\overline{\Psi}\gamma^i\Psi,
\end{split}
&
\begin{split}
V^0&= \overline{\Psi}\gamma^0\Psi,\\
A^i&= \overline{\Psi}\gamma_5\gamma^i\Psi,
\end{split}
&
\begin{split}
A^0&=\overline{\Psi}\gamma_5\gamma^0\Psi,\\
Q^i&=\overline{\Psi}\gamma^0\gamma_5\gamma^i\Psi.
\end{split}\label{Eq_Bilin_0}
\end{align}

To obtain dynamical equations for each of these bilinears we consider projections of the Dirac equation of the following form:
\begin{align}
i\Psi^\dagger M \frac{\delta W}{\delta \overline{\Psi}} \pm i\left(\frac{\delta W}{\delta \overline{\Psi}}\right)^\dagger M\Psi,\label{projections}
\end{align}
where $M = \{\mathbb{I},\gamma_5,\gamma^0,\gamma_5\gamma^0,\gamma^i,\gamma_5\gamma^i,\gamma^0\gamma^i,\gamma_5\gamma^0\gamma^i\}$ is one of 8 possibilities taken from the four dimensional Clifford algebra. Through this procedure we obtain the following eight projected Dirac equations,
\begin{subequations}
\begin{align}
\dot{E}&= -3HE  +4\xi B A^0,\\
\dot{B}&= - 3
HB -2(U'+2\xi E)A^0,\\
\dot{A}^0 &=- 3HA^0 +2(U' + 2\xi E)B - 4\xi BE,\\
\dot{V}^0
&=-3HV^0 ,  \\
\dot{C}^i 
&=-3HC^i +2(U'+2\xi E)V^i,\\
\dot{Q}^i&=-3HQ^i+4\xi BV^i,\\
\dot{V}^i
&=
-3H V^i-2(U'+2\xi E)C^I
-4\xi BQ^i,\\
\dot{A}^i
&= - 3HA^i\,.
\end{align}\label{eq_ambi_background}
\end{subequations}

\subsection{Parity invariant bouncing solutions}
\label{Parityinv}
Having chosen our flat FLRW background ansatz for the metric, we are now in a position to obtain analytic solutions to the equations of motion. Given the form of the metric, it is reasonable - as well as computationally advantagous - to consider solutions in which the background spinor is also parity invariant (although for a more complete discussion of parity violations in these models the reader should consult~\cite{Freidel:2005sn}). That is, we consider the `ambidextrous' case of~\cite{Magueijo:2012ug}, in which the background spinors satisfy:
\begin{align}
\gamma^0\Psi = \Psi.
\end{align}
For parity invariant spinors, bilinears which sandwich an odd number of spatial gamma matrices will always be zero. For example $\overline{\Psi}\gamma^i\gamma_5\gamma^j\Psi = 0$. This implies $V^i = C^i = A^0 = B = 0$, while  $Q^i = A^i$, and $E = V^0$. The equations of motion therefore simplify further, and only four remain which will be of relevance to us:
\begin{subequations}
\begin{align}
\dot{E} + 3HE&=
0,\label{Eq_Back_Fin_E}\\
\dot{A}^i+3HA^i&=0,\label{Eq_Back_Fin_A}\\
12\kappa H^2 &= [U + \xi E^2],\label{Eq_Back_Fin_GR00}\\
-4\kappa(2\dot{H} + 3H^2)&=(\xi E^2 + U'E - U),\label{Eq_Back_Fin_GRij}
\end{align}
\label{BackgroundVmE}
\end{subequations}
It is possible to directly solve the projected Dirac equations given in Eq.~\eqref{Eq_Back_Fin_E} and~\eqref{Eq_Back_Fin_A}, yielding the results:
\begin{align}
E &= \frac{M}{a^3}, & A^i &= \frac{\alpha^i}{a^3},\label{Eq_Back_E}
\end{align}
where, following Eq.~\eqref{Eq_Bilinearex_1}, $M,\alpha^i$ are time independent constants satisfying $M^2 = \alpha^i\alpha_i$. Interestingly this result is true for any time dependence of the background
geometry (as noticed before~\cite{Magueijo:2012ug}), and so the spinor density and axial current monotonically increase in a contracting universe.

We would also like to solve Eq.~\eqref{Eq_Back_Fin_GR00} to obtain a background solution for the scale factor. However such a solution will necessarily depend on the choice of potential $U$. Fortunately, given our solution for the spinor density $E$, it is very easy to select the potential $U$ such that the $\{0,0\}$ component of the Einstein equations takes the following tractable form:
\begin{align}
\dot{a}^2 = \frac{c_1}{a^{3n-2}} + \frac{c_2}{a^{6n-2}}, \label{Eq_diff_anform}
\end{align}
where $c_1$ and $c_2$ are time independent constants. Equation~\eqref{Eq_diff_anform} has the following solution:
\begin{align}
a(t) = (-\tfrac{c_2}{c_1}+\tfrac{9}{4}c_1n^2t^2)^{1/3n}.
\end{align}
which undergos a bounce rather generically so  long as $c_1$ is positive and $c_2$ is negative. As an example consider the case in which the potential is simply given by a mass term for the spinor $U = m E$. In this case the  $\{0,0\}$ component of the Einstein equation is precisely of the form given in  Eq.~\eqref{Eq_diff_anform} for  $n = 1$. This is the so-called `borderline' scenario found in~\cite{Magueijo:2012ug}, in which the matter density scales in the same way as the anisotropies during a contracting phase. In this case the solution for the scale factor is given by
\begin{align}
U &= mE, & a(t) &= \left[M(-\frac{\xi}{m} + \frac{3m t^2}{16\kappa})\right]^{1/3}\,,\label{Eq_back_asol}
\end{align}
with $\xi < 0.$ While this solution is interesting we would like to see how easy it is to obtain not only a bouncing solution, but also to control the dynamics away from the bounce. For example, is it possible to obtain a phase of inflation following the bounce, or a period of ekpyrosis prior? Spinor inflaton fields have already been discussed elsewhere in the literature (see eg.~\cite{ArmendarizPicon:2003qk}), and so for the sake of interest we will consider the example of an ekpyrotic model.

\begin{figure}[h]
	\begin{minipage}{0.5\textwidth}
	\includegraphics[width=0.9\textwidth]{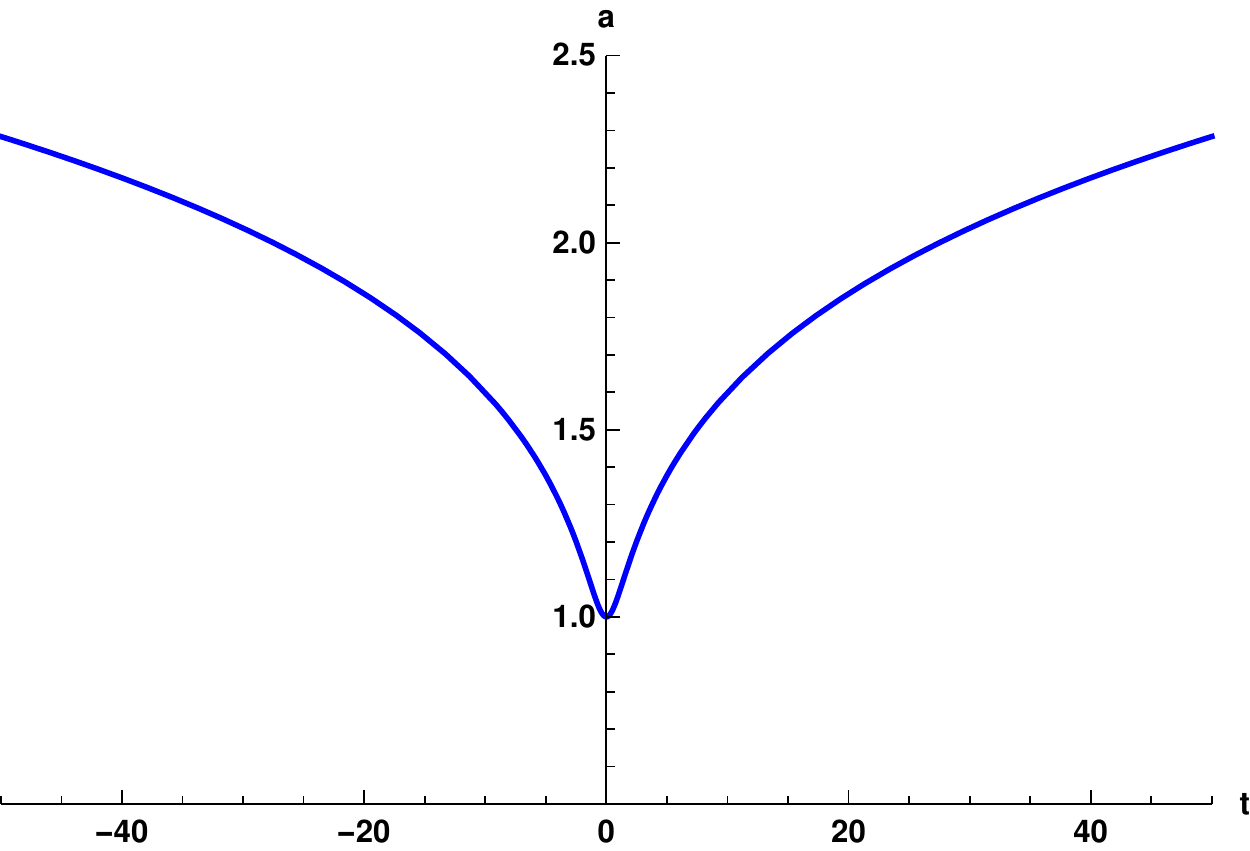}
\end{minipage}%
\begin{minipage}{0.5\textwidth}
	\includegraphics[width=0.9\textwidth]{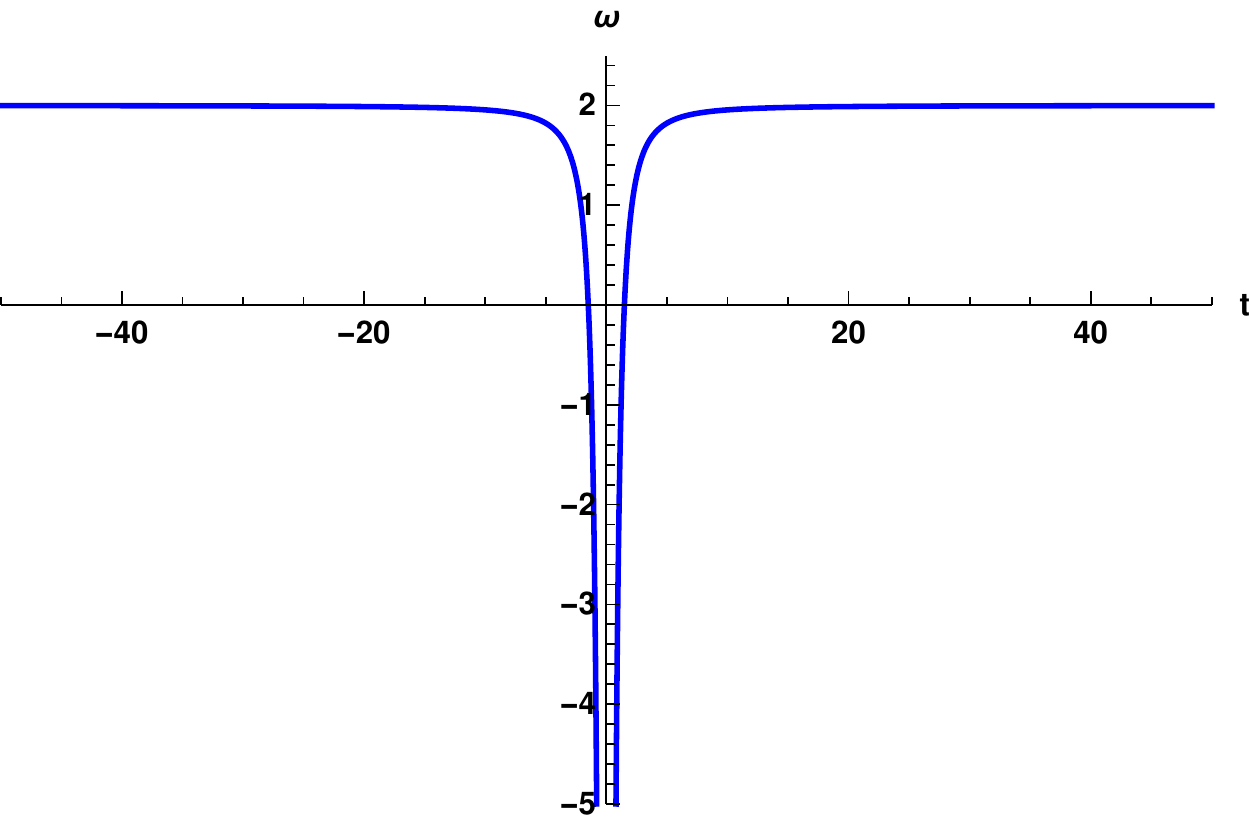}
	\end{minipage}%
	\caption
	{Evolution of the scale factor $a$ (left panel)  and the equation of state $\omega$ (right panel) for choice of potential $U = -\xi E^2 +b_1 E^3 +b_2 E^{6}$, with $b_1 = 0.1$, $b_2 = -0.1$,  $M=1$, and $\kappa = 1/4$. In this model an ekpyrotic contraction phase is followed by a non-singular bounce into an expanding phase. The ekpyrotic phase renders the universe flat and isotropic in the approach to the bounce, and justifies the assumption of a flat FLRW metric in describing the bounce.}
	\label{fig:bounce}
\end{figure}

In some sense, having an ekpyrotic phase before the bounce is not really optional, but rather necessary: in a contracting universe small anisotropies grow as $a^{-6}$ and, in the absence of a faster-growing energy component, the anisotropies quickly come to dominate the dynamics, thus preventing a smooth non-singular bounce from occurring. Thus, if we want to explain the required isotropy of the contracting universe just prior to the bounce in a dynamical fashion, we need an ekpyrotic phase. The stiff equation of state $P> \rho$ during ekpyrosis suppresses anisotropies and renders the universe flat and smooth in the approach to the bounce \cite{Erickson:2003zm,Lehners:2008vx}. Moreover, some models of ekpyrosis can generate the density perturbations seen as temperature fluctuations in the cosmic microwave background (see e.g. \cite{Lehners:2013cka,Ijjas:2015hcc} and references therein). To implement an ekpyrotic phase, followed by a bounce, consider the following potential:
\begin{align}
U(E) = -\xi E^2 +b_1 E^n + b_2 E^{2n},
\end{align}
for integer $n$. Now the Einstein equation is once again of the desired form given in Eq.~\eqref{Eq_diff_anform}, while the equation of state is given by
\begin{align}
\omega = \frac{P}{\rho} = \frac{ U'E - U+ \xi E^2}{U+\xi E^2} =
(n-1)+\frac{nb_2 E^{2n}}{b_1 E^n+b_2 E^{2n}}
 \end{align}
Because $E$ decreases monotonically with growing scale factor, the equation of state approaches $(n-1)$ far away from the bounce, while becoming negative (and of large magnitude) as the bounce is approached. For an ekpyrotic phase we require $\omega > 1$, which means that we need to take $n > 2$. For $n=3$ we can solve the $\{0,0\}$ component of the Einstein equations exactly to obtain the following solution for the scale factor:
\begin{align}
a(t) = \left[M^3\left(-\frac{b_2}{b_1}+\frac{27 b_1 t^2 }{16\kappa}\right)\right]^{1/9}.
\end{align}
As shown in Fig.~\ref{fig:bounce}, this solution neatly combines an ekpyrotic contracting phase with a cosmological bounce leading into an expanding phase of the universe.  

\subsection{Flat FLRW and anti-commuting spinors}

In the previous subsections we considered background solutions  for commuting spinors satisfying the identities given in Eq.~\eqref{Eq_Bilinearex_1}. While commuting spinors are simpler to work with computationally, they do not correspond to any of the fermions known in the standard model. In this section we work with anti-commuting spinors. We once again ask that the quartic spinor terms in the equations of motion may be taken to be of the form $\langle A_I\rangle \langle A^I\rangle$. This can be achieved either by assuming the variance of spinor bilinears is low, or by assuming that the first order description is fundamental. Given either of these assumptions the Einstein Equations for a flat FLRW background are given by:
\begin{subequations}
\begin{align}
12\kappa H^2 &=  W\\
-4\kappa(2\dot{H}+3H^2)&= (W +U'E - 2U)
\end{align}
where the potential $W$ is given as in Eq.~\eqref{FullMod_EOM_omega_5}. The projected Dirac equations are given by:
\begin{align}
\dot{E}&=-3HE + 4(\xi_{AA}A^0 + \xi_{AV}V^0)B + 4(\xi_{VV}V^i +\xi_{AV}A^i)C_i,\\
\dot{B} &=-3HB-4(\xi_{AA}A^0+\xi_{AV}V^0)E +
4(\xi_{VV}V^i+\xi_{AV}A^i)Q_i,\\
\dot{A}^0 &=-3HA^0 + 2U'B,\\
\dot{V}^0&= -3HV^0,\\
\begin{split}
\dot{C}^i&=-3HC^i + 2U'V^i+ 4(\xi_{AA}A_j + \xi_{AV}V_j)\varepsilon^{0jik}C_k\\
&~~~- 4[(\xi_{AA}A_0+\xi_{AV}V_0)Q^i+ (\xi_{VV}V^i +\xi_{AV}A^i)E], \end{split}\\
\dot{Q}^i &= -3HQ^i +4(\xi_{AA}A_0+\xi_{AV}V_0)C^i + 4(\xi_{VV}V^i +\xi_{AV}A^i)B-4\varepsilon^{0ijk}Q_j(\xi_{AA}A_k+\xi_{AV}V_k)\\
\dot{V}^i &= -3HV^i - 2 U'C^i +4[\xi_{AA}A_j\varepsilon^{0jik}V_k + \xi_{VV}V_j \varepsilon^{0jik}A_k]\\
\dot{A}^i&=-3HA^i.
\end{align}
\end{subequations}
We are not able to solve the above equations analytically, even when restricting attention to parity invariant anti-commuting spinors. This does not prevent us from making progress however, as we can solve the above equations numerically. We consider the simple case in which the couplings $\xi_{AV}$ and $\xi_{VV}$ are both set to zero, and for which the potential is given by a mass term for the spinor: $U = mE$. The projected Dirac equations then simplify considerably. We plot solutions for these equations for a particular choice of initial conditions below in figures \ref{fig:bounce-anti} and \ref{fig:bounce-density-anti}. Once again a non-singular bouncing solution is obtained. In fact, in the solution shown in the figures the null energy condition is violated twice, leading to a ``double'' bounce. Note that in the anti-commuting case bounces arise for $\xi_{AA} > 0,$ i.e. for the opposite sign of the coupling than in the commuting case. Also, the axial vector is timelike instead of spacelike for these solutions. 

The numerical solutions we have obtained provide a `proof of principle', that commuting spinnors are not a necessary requirement  in order to obtain  bouncing solutions. It will certainly be interesting to investigate the properties of anti-commuting spinor bounces in more detail, to see how general they are. Here we simply note the evident similarity with the commuting case. For now we will return to the technically simpler case of commuting spinors, in order to assess the stability of such non-singular bouncing solutions.
\begin{figure}[h]
	\begin{minipage}{0.5\textwidth}
	\includegraphics[width=0.9\textwidth]{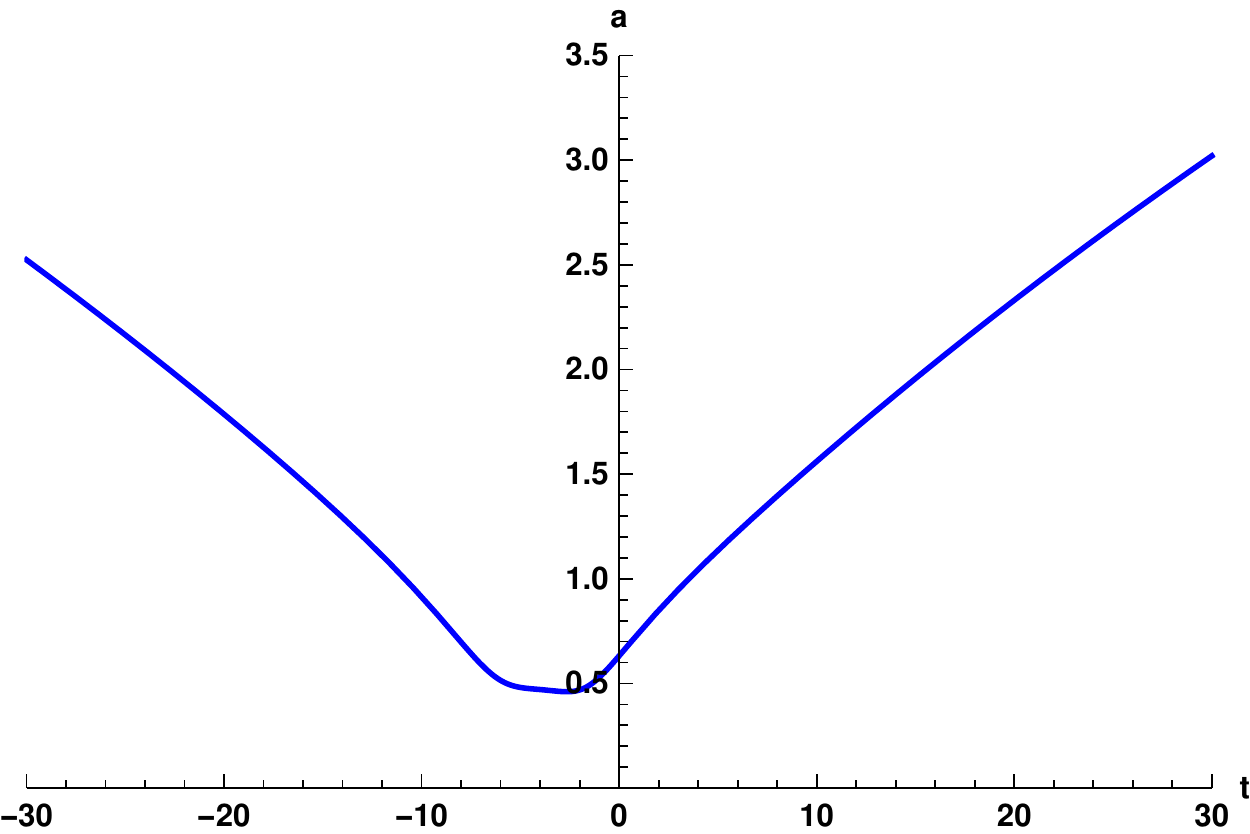}
\end{minipage}%
\begin{minipage}{0.5\textwidth}
	\includegraphics[width=0.9\textwidth]{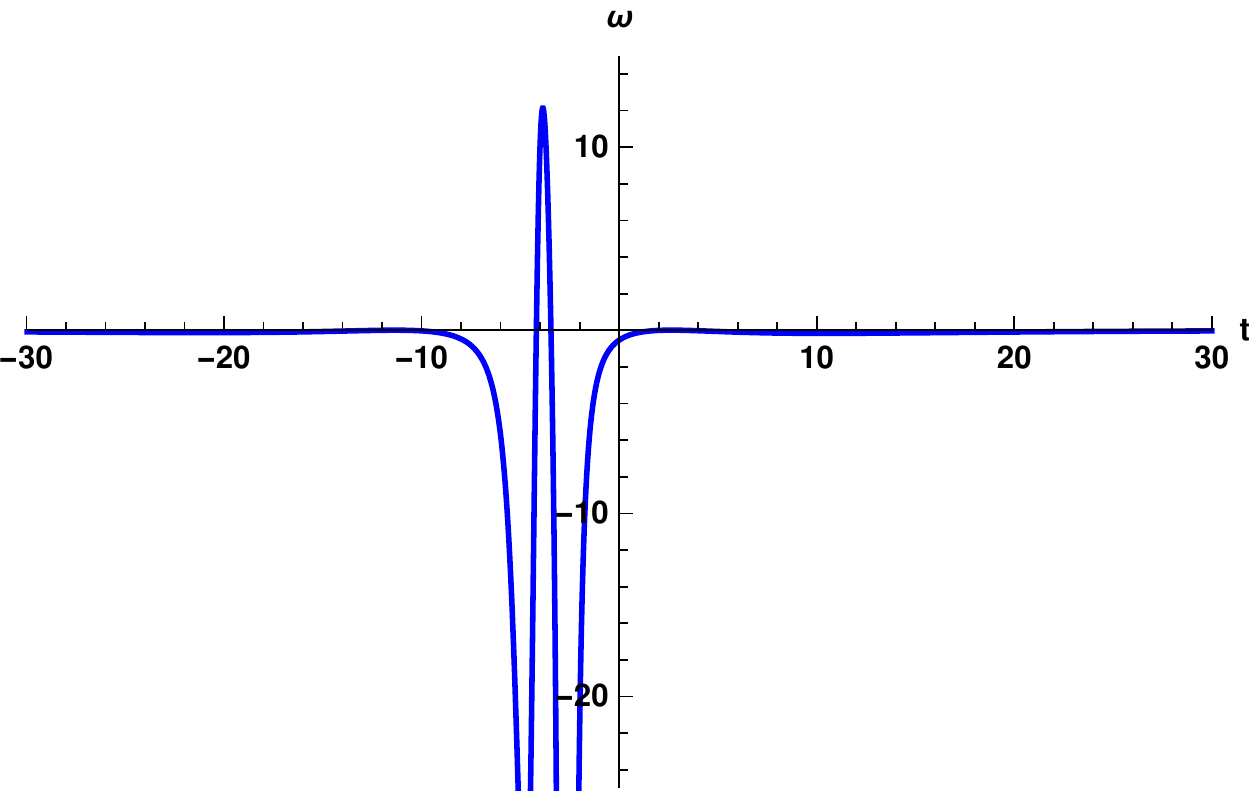}
	\end{minipage}%
	\caption
	{Evolution of the scale factor $a$ (left panel)  and the equation of state $\omega$ (right panel) for a particular set of initial conditions, and the choice of potential $U = m E$, with $m = 0.5$, $M=1$, $\kappa = 1/4$, and choice of couplings $\xi_{AA} = 0.15$, $\xi_{VA} = \xi_{VV} = 0$.}
	\label{fig:bounce-anti}
\end{figure}

\begin{figure}[h]
	\begin{minipage}{0.5\textwidth}
	\includegraphics[width=0.9\textwidth]{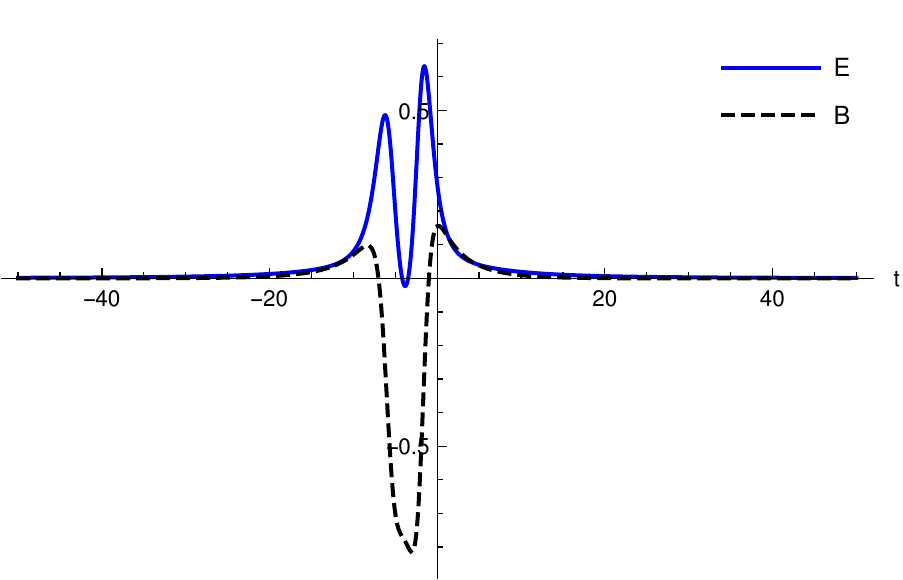}
\end{minipage}%
\begin{minipage}{0.5\textwidth}
	\includegraphics[width=0.9\textwidth]{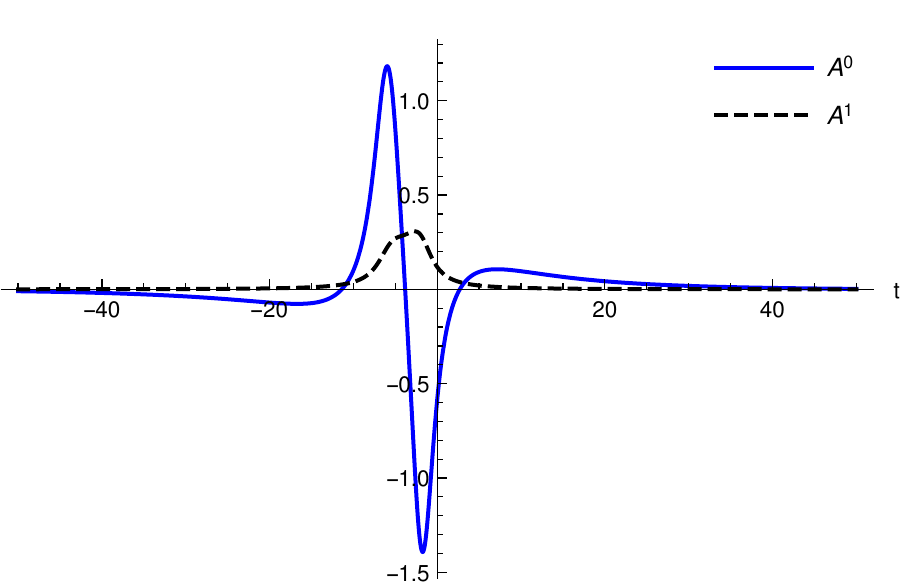}
	\end{minipage}%
	\caption
	{Evolution of the spinor densities $E$ and $B$ (left panel) as well as the axial vector components $A^0$ and $A^1$ (right panel) for a particular set of initial conditions, and the choice of potential $U = m E$, with $m = 0.5$, $M=1$, $\kappa = 1/4$, and choice of couplings $\xi_{AA} = 0.15$, $\xi_{VA} = \xi_{VV} = 0$. }
	\label{fig:bounce-density-anti}
\end{figure}

\section{Perturbing around flat FLRW}
\label{Sec_Perturbations}

So far we have treated the universe as perfectly homogeneous and isotropic. We will now introduce inhomogeneities by perturbing around our flat FLRW background solutions in order to address the questions of stability and of observational consequences. In our preliminary foray we return to the case of  commuting spinors for the purpose of computational simplicity. In future work we plan to make a more complete analysis which includes anti-commuting spinors.

\subsection{Linearised equations of motion in Newtonian gauge}

In this section we analyse the linearised equations of motion using a standard Fourier decomposition. We work in Newtonian gauge, and provide a complete description of our gauge fixing procedure in the appendix for both the metric and vierbein perturbations. The perturbed FLRW line element in Newtonian gauge is expressed as
\begin{align}
ds^{2}=-(1+2\psi)d\tau^{2}+2a(t)B_{i}d\tau dx^{i}+a(t)^{2}((1-2\phi)\delta_{ij}+h_{ij})dx^{i}dx^{j}~,\label{lineelement}
\end{align}
where $\psi$, $B_{i}$ and $h_{ij}$ are perturbations which depend a priori on all spacetime coordinates, and where $\partial^{i}B_{i}=h_{i}^{i}=\partial^{i}h_{ij}=0$. For the perturbed Dirac spinor $\Psi$ we introduce the following notation: 
\begin{align}
\Psi=	\Psi_{(0)}+\Psi_{(1)}~,
\end{align}
where the subscripts $(0)$ and $(1)$ label background and first order quantities respectively. Although our background spinor solutions $\Psi_{(0)}$ are spatially independent, we allow for general spatial dependence of the perturbation $\Psi_{(1)}$. Given this notation, the first order momentum space Dirac equation is:
\begin{align}
\begin{split}
(1 - \psi)\gamma^0  \dot{\Psi}+ i\gamma^i  \tfrac{k_i}{a} \Psi
 &= - \tfrac{3}{2} \gamma^0 [H-\dot{\phi}-H\psi]\Psi
 -i(U'(E)\Psi  + 2\xi(E\Psi - iB\gamma_5\Psi))
\\
&~~~ -\tfrac{1}{2}\gamma^i[i\tfrac{k_i}{a}\psi - 2i\tfrac{k_i}{a}\phi +\tfrac{1}{2}\dot{B}_i]\Psi_{(0)}
 +\tfrac{i}{2}(U'+2\xi E)B_i\gamma^i\Psi_{(0)},
 \end{split}\label{Eq_first_Dir}
\end{align}
where for compactness we have included some background terms, and some terms higher than first order. Any terms that are not of order 1 should be ignored by the reader. For example the left hand side of Eq.~\eqref{Eq_first_Dir} is intended to be read as: $- \psi  \partial_0 \Psi_{(0)}+\gamma^0  \partial_0 \Psi_{(1)}+ i\gamma^i  k_i \Psi_{(1)}$. We adopt this compact notation often throughout the remainder of the section.

In Section~\ref{Sec_Background_FLRW}, we found it useful to express the background Dirac equation in projected form, written  entirely in terms of spinor bilinears. Following an analogous procedure, we similarly project the linearized Dirac equation. Because we are interested in parity invariant  background solutions satisfying $\Psi_{(0)} = \gamma^0\Psi_{(0)}$, there are only eight possible `kinds' of first order hermitian bilinears which may be constructed from the background spinor $\Psi_{(0)}$ and its perturbation $\Psi_{(1)}$. These are:
\begin{align}\label{Eq_Bilin_1}
\begin{split}
E_{(1)} &=(\overline{\Psi}_{(0)}\Psi_{(1)}+ \overline{\Psi}_{(1)}\Psi_{(0)}),\\
A^0_{(1)}&=(\overline{\Psi}_{(0)}\gamma_5\gamma^0
\Psi_{(1)} 
+\overline{\Psi}_{(1)}\gamma_5\gamma^0\Psi_{(0)}),\\
V^i_{(1)} &=(\overline{\Psi}_{(0)}\gamma^i\Psi_{(1)}+
\overline{\Psi}_{(1)}\gamma^i\Psi_{(0)}),\\
A^i_{(1)}&= (\overline{\Psi}_{(0)}\gamma_5\gamma^i\Psi_{(1)}+
\overline{\Psi}_{(1)}\gamma_5\gamma^i\Psi_{(0)}),
\end{split} 
&
\begin{split}
i\widetilde{E}_{(1)} &=(\overline{\Psi}_{(0)}\Psi_{(1)}- \overline{\Psi}_{(1)}\Psi_{(0)}),\\
iB_{(1)} &= (\overline{\Psi}_{(0)}\gamma_5\Psi_{(1)} + 
\overline{\Psi}_{(1)}\gamma_5\Psi_{(0)}),\\
iC^i_{(1)} &= (\overline{\Psi}_{(0)}\gamma^i\Psi_{(1)}
-\overline{\Psi}_{(1)}\gamma^i\Psi_{(0)}), \\
i\widetilde{A}^i_{(1)}&=(\overline{\Psi}_{(0)}\gamma_5
\gamma^i\Psi_{(1)}-\overline{\Psi}_{(1)}\gamma_5
\gamma^i\Psi_{(0)}),
\end{split}
\end{align}
where our naming convention corresponds as closely as possible with that of the background bilinears defined by Magueijo \textit{et al.}~\cite{Magueijo:2012ug}. We point out in particular that the axial scalar $B_{(1)}$ should not be confused with the metric vector purturbation $B^i$, which always apears with an index. In projected form, the linearized Dirac equations are expressed in terms of these eight `kinds' of spinor bi-linears as:
\begin{subequations}
\begin{align}
\dot{E}_{(1)}&= -3HE_{(1)} +3\dot{\phi}E_{(0)}- i\tfrac{k_i}{a} V^i_{(1)}  \label{Eq_first_Dir_E},\\
\dot{A}^j_{(1)}
&= -3HA^j_{(1)}+3\dot{\phi}A^j_{(0)}
-i\tfrac{k^j}{a}A^0_{(1)}-i\varepsilon^{0ji}_{~~~k}
\tfrac{k_i}{a}C^{k}_{(1)},\label{Eq_first_Dir_AI}\\
\dot{B}_{(1)}&=
-3
HB_{(1)}+i\tfrac{k_i}{a}\widetilde{A}^i_{(1)} -2(U_{(0)}'+2\xi E_{(0)})A^0_{(1)}-(U'_{(0)} + 2\xi E_{(0)})B_iA^i_{(0)},\label{Eq_first_Dir_B}\\
\dot{A}^0_{(1)}&=- 3H A^0_{(1)}+2U_{(0)}'B_{(1)}   -
i\tfrac{k_i}{a} A^i_{(1)}-[i\tfrac{k_i}{a}\psi - 2i\tfrac{k_i}{a}\phi +\tfrac{1}{2}\dot{B}_i]A^i_{(0)},\label{Eq_first_Dir_A0}\\
\begin{split}
\dot{C}^j_{(1)}&=-3HC^j_{(1)}+ 2(U_{(0)}'+2\xi E_{(0)}) V^j_{(1)}
-i\tfrac{k^j}{a} \widetilde{E}_{(1)}+i\varepsilon^{0ji}_{~~~k}\tfrac{k_i}{a}A^{k}_{(1)}\\
&~~~+\varepsilon^{0ji}_{~~~k}[i\tfrac{k_i}{a}\psi - 2i\tfrac{k_i}{a}\phi +\tfrac{1}{2}\dot{B}_i]A^k_{(0)}
+(U_{(0)}'E_{(0)}+2\xi E^2_{(0)})B^j,\end{split}\label{Eq_first_Dir_C}\\
\begin{split}
\dot{V}^j_{(1)}&=-3H V^j_{(1)}
-i\tfrac{k^j}{a} E_{(1)}
-i\varepsilon^{0ji}_{~~~k}\tfrac{k_i}{a}\widetilde{A}_{(1)}^{k}-2\left[(U'_{(0)}+2\xi E_{(0)})C^j_{(1)}
+2\xi B_{(1)}A^j_{(0)}\right]
\\
&~~~-[i\tfrac{k^j}{a}\psi - 2i\tfrac{k^j}{a}\phi +\tfrac{1}{2}\dot{B}^j]E_{(0)}
+(U'_{(0)}+2\xi E_{(0)})B_i\varepsilon^{0ji}_{~~~k}A^k_{(0)}
\label{Eq_first_Dir_V},\end{split}\\
\begin{split}
\dot{\widetilde{E}}_{(1)}&=-3H \widetilde{E}_{(1)} -i\tfrac{k_i}{a}C^i_{(1)} \\
&~~~-2(U_{(1)}' + 2\xi E_{(1)})E_{(0)} - 2\psi(U'_{(0)} + 2\xi E_{(0)})E_{(0)}\label{Eq_first_Dir_ET},\end{split}\\
\begin{split}
\dot{\widetilde{A}^i}_{(1)}&=
-3H\widetilde{A}^i_{(1)}  + i\varepsilon^{0ij}_{~~~k}\tfrac{k_j}{a}V^{k}_{(1)}+i\tfrac{k^i}{a}B\\
&~~~-2(U'_{(1)} + 2\xi E_{(1)})A_{(0)}^i-2\psi(U_{(0)}'+2\xi E_{(0)})A_{(0)}^i\label{Eq_first_Dir_AT},\end{split}
\end{align}\label{Eq_first_Dir_Proj}
\end{subequations}
In component form, there are $16$ spinor bilinears, and correspondingly $16$ projected Dirac equations. We remind the reader that the spinor bilinears do not all represent independent degrees of freedom as they derive from the same spinor $\Psi$. For example, we know that for commuting spinors, the bilinears $E,B,V^I,$ and $A^I$ are related by the identity given in Eq.~\eqref{Eq_Bilinearex_1}. Further relationships can be found between the various bilinears by use of the Fierz identity given in Eq.~\eqref{Id_Fierz}. 

After some manipulation the first order Einstein equations can be written in the following compact form, in which the stress energy tensor is expressed entirely in terms of spinor bilinears:
\begin{subequations}
\begin{align}
\begin{split}
4\kappa G_{00} &= -8\kappa[\tfrac{k_ik^i}{a^2}\phi + 3H\dot{\phi}]\\
&=
(1+2\psi)(U+  \xi E^2)+\tfrac{i}{4}\tfrac{1}{a}[B_{[i}k_{k]}\varepsilon^{0ikj}A_j+2k_i C^i],  
\end{split}\label{Eq_first_Eins00}\\
4\kappa G_{i0} &= 4\kappa[2i\tfrac{k_i}{a}(\dot{\phi} + H\psi) -(2\dot{H}+3H^2)B_i + \tfrac{1}{2}\tfrac{k_kk^k}{a^2}B_i]\nonumber\\
\begin{split}
&=-\tfrac{1}{4}[2i\tfrac{k_i}{a}\widetilde{E}
  +\eta_{in}i\varepsilon^{0mnk}A_{k}\tfrac{k_m}{a}
-i(3\eta_{in}\phi \tfrac{k_m}{a} +
 \tfrac{1}{2}h_{im}\tfrac{k_n}{a})\varepsilon^{0mnk}A_k]\\
& ~~~~~~~~~+ B_i(\xi E^2 + U'E - U),\end{split}\label{Eq_first_Einsi0}\\
4\kappa G_{ij} &= 4\kappa\Big[[-\tfrac{k^kk_k}{a^2}(\psi-\phi) + 2\ddot{\phi} + 2(2\dot{H}+ 3H^2)(\phi+\psi) + 2H\dot{\psi} + 6H\dot{\phi}]\delta_{ij}-\tfrac{k_ik_j}{a^2}(\phi-\psi)
\nonumber\\
&~~~~~~-i\tfrac{1}{a}(\dot{B}_{(i}k_{j)}+2HB_{(i}k_{j)})+\tfrac{1}{2}\ddot{h}_{ij}+\tfrac{1}{2}\tfrac{k^kk_k}{a^2} h_{ji}+ \tfrac{3}{2}H\dot{h}_{ij} 
-(2\dot{H}+ 3H^2)h_{ij}\Big]\nonumber\\
\begin{split}
&=\tfrac{1}{4}\Big[2i\tfrac{1}{a}C_{(i}k_{j)}-\tfrac{1}{2}[(\dot{h}_{jl}-iB_{j}\tfrac{k_{l}}{a})\eta_{ik}+ (\dot{h}_{il}-iB_{i}\tfrac{k_{l}}{a})\eta_{jk}]\varepsilon^{kl0m}A_m\Big]\\
&~~~ +((1-2\phi)\eta_{ij} +h_{ij})[\xi E^2 + U'E - U]
 .\end{split}\label{Eq_first_Einsij}
\end{align}\label{Eq_first_Eins}
\end{subequations}
The standard approach when perturbing about FLRW is to separate the equations of motion at linear order into their scalar, vector, and tensor components. This procedure is known as the SVT decomposition and greatly simplifies the analysis. Unfortunately, we are not able to follow the standard approach here, as can be seen for example by considering the $B_{[i,k]}\varepsilon^{0ikj}A_j$  term present in Eq.~\eqref{Eq_first_Eins00}. This term is a scalar in the sense that all of its indices are fully contracted, however it is clearly built from the vector perturbation $B_i$. Similar couplings between scalar, vector, and tensor modes can be seen in Eqs.~\eqref{Eq_first_Einsi0} and~\eqref{Eq_first_Einsij}, as well as in the first order projected Dirac equations. 

To understand this point, notice that the proof of the SVT decomposition theorem is highly dependent on the symmetries of the background, and requires that no relevant background quantity can be formed which violates this symmetry (see for example the appendix of~\cite{Baumann:2009ds}). As seen in Eq.~\eqref{Eq_Bilinearex_1} however, the models we consider all fail this requirement explicitly because the background axial vector $A_{(0)}^i$ picks out a preferred spacelike direction. In practice this allows first order terms to be constructed in which the index on a perturbation is contracted with the index on $A_{(0)}^i$, rather than always having to transform as a free index, or else contract with non-symmetry-breaking projectors such as $k_i$. This is precisely the kind of SVT mixing that we observe in the equations of motion.

Despite the difficulty of not being able to completely decompose the equations of motions into separate scalar, vector, and tensor parts, this does not prevent us from making considerable progress. For example, we are able to make simplifications by considering contractions of the $\{0,i\}$ Einstein equation with $k^i$, and also the $\{i,j\}$ Einstein equation with $k^i k^j$ and $\eta^{ij}$. After some manipulations this procedure yields the following scalar equations:
\begin{subequations}
\begin{align}
\widetilde{E}&=-16\kappa (\dot{\phi} + H\psi),\\
8\kappa
\left( \ddot{\phi} + (2\dot{H}+ 3H^2)\psi + H(\dot{\psi} + 3\dot{\phi}) \right)
&=\tfrac{i}{2}\tfrac{k^j}{a}C_j+(\xi E^2 + \frac{\partial U}{\partial E}E - U)],\\
4\kappa (\phi-\psi) 
&=\tfrac{i}{8k^kk_k}a(k_lB_{k}\varepsilon^{kl0m}A_m-4k_i C^i).
\end{align}\label{Kdata} 
\end{subequations}
Substituting these scalar relations back into the first order Einstein equations, we can also obtain simplified equations for the vector and tensor modes:
\begin{subequations}
\begin{align}
8\kappa \tfrac{k_kk^k}{a^2}B_i&=i[\eta_{in}\varepsilon^{0mnk}A_{k}\tfrac{k_{m}}{a}
-(3\eta_{in}\phi \tfrac{k_{m}}{a} +
 \tfrac{1}{2}h_{im}\tfrac{k_n}{a})\varepsilon^{0mnk}A_k], \label{Eq_first_eini0new} \\
\begin{split}
-8\kappa \tfrac{k_ik_j}{a^2}(\phi-\psi)
 &= \Big[i\tfrac{1}{a}C_{(i}k_{j)}-\tfrac{1}{4}[iB_{k}\tfrac{k_{l}}{a}\delta_{ij}+(\dot{h}_{jl}-iB_{j}\tfrac{k_{l}}{a})\eta_{ik}+ (\dot{h}_{il}-iB_{i}\tfrac{k_{l}}{a})\eta_{jk}]\varepsilon^{kl0m}A_m\Big]\\
 &~~~+4\kappa\Big[
i\tfrac{1}{a}(\dot{B}_{(i}k_{j)}+2HB_{(i}k_{j)})
-\tfrac{1}{2}\ddot{h}_{ij}-\tfrac{1}{2}\tfrac{k^kk_k}{a^2} h_{ji}- \tfrac{3}{2}H\dot{h}_{ij} 
\Big]. \end{split}\label{Eq_first_einijnew}
\end{align} \label{Eq_first_einijnewall}
\end{subequations}

\subsubsection{Real fluid description}

Before attempting to solve the linearized equations of motion, we note that it is conceptually useful to re-express the equations of motion  in  fluid form.  Unlike for the highly symmetric background, the first order contribution to the stress energy tensor can not be expressed as a perfect fluid, but instead take the more general form of a real fluid:
\begin{align}
T^\mu_{~\nu} = u^\mu u_\nu(P+\rho) + \delta^\mu_\nu P + \Sigma^\mu_{~\nu},
\end{align}
where $\Sigma_{\mu\nu}$ is the anisotropic stress satisfying:
\begin{align}
\Sigma_{\mu\nu}&=\Sigma_{\nu\mu}, & \Sigma_{\mu\nu}u^\nu &=0, & \Sigma_{\mu}^{\mu}=0,\label{anisotropprop}
\end{align}
with $u_{\mu} = \{-(1+\psi),av_i\}$, $u^\mu = \{(1-\psi),\tfrac{1}{a}(v^i-B^i)\}$, and where $v_i$ is the peculiar velocity of the fluid. Making use of the relations given in Eq.~\eqref{anisotropprop}, together with equations~\eqref{Eq_first_Eins}, we determine the first order contributions to the density $\rho$, pressure $P$ and anisotropic stress $\Sigma_{ij}$ to be:
\begin{subequations}
\begin{align}
\rho &=(U +\xi E^2)+\tfrac{i}{4a}[B_{[i}k_{k]}\varepsilon^{0ikj}A_j+2k_k C^k],\\
P
&=[\xi E^2 + U'E - U]  + \tfrac{i}{12a}[B_{k}k_l\varepsilon^{kl0m}A_m +2k_k C^k],\\
\Sigma_{ij} &=\tfrac{i}{4}\Big[2\tfrac{1}{a}C_{(i}k_{j)}-\tfrac{2}{3}\eta_{ij}\tfrac{k_k}{a}C^k +\tfrac{1}{2}[(i\dot{h}_{jl}+B_{j}\tfrac{k_l}{a})\eta_{ik}+ (i\dot{h}_{il}+B_{i}\tfrac{k_l}{a})\eta_{jk}-\tfrac{2}{3}\eta_{ij}B_{k}\tfrac{k_l}{a}]\varepsilon^{kl0m}A_m\Big],
\end{align}
where we are once again using compact notation, in which we have included background terms, and terms higher than first order. It is also useful to define the 3-momentum density:
\begin{align}
q_i\equiv (\rho + P)v_i&=
\tfrac{i}{4a}[2k_i\widetilde{E}
-((1-3\phi)\eta_{in} +
 \tfrac{1}{2}h_{in})k_m\varepsilon^{0mnk}A_k].
\end{align}
The scalar components of the anisotropic stress tensor, and 3-momentum density are given respectively by:
\begin{align}
\Sigma &=
\tfrac{i}{8k^kk_k}a(k_lB_{k}\varepsilon^{kl0m}A_m-4k_i C^i),\\
q 
&= \tfrac{1}{2}\widetilde{E}. \label{qandE}
\end{align}\label{Eq_first_fluiddef}
\end{subequations}
Here we see that anisotropic stress is induced both from the vector perturbations $B_i$, as well as  the spinor bilinear $C^i_{(1)}$.   
Using the definitions given above for $\rho,P,\Sigma,$ and $q$, equations~~\eqref{Eq_first_Eins00} and~\eqref{Kdata} can be written in the standard real fluid form
\begin{subequations}
\begin{align}
4\kappa (\phi-\psi) 
&=\Sigma ,\\
8\kappa[ \tfrac{k^2}{a^2}\phi + 3H\dot{\phi}+3H^2\psi]
&= 
-\rho,\label{diffthis}
\\
8\kappa (\dot{\phi} + H\psi)
&=-q, \label{linearq} \\
8\kappa[ \ddot{\phi} + (2\dot{H}+ 3H^2)\psi + H(\dot{\psi} + 3\dot{\phi})]
&=P - \tfrac{2}{3}\tfrac{k^2}{a^2}\Sigma.
\end{align}
We may also use the projected Dirac equation to derive evolution equations, including the equation of continuity:
\begin{align}
 \dot{\rho} &=-3(H-\dot{\phi})(\rho+P)+\tfrac{k^2}{a^2} q, \\
 \dot{q}&=-3Hq
-\psi(P+\rho)-P + \tfrac{2}{3a^2}k^ik_i\Sigma,\\
\begin{split}
\dot{P} &= -6[H-\dot{\phi}](P + U - U'E + \tfrac{1}{2}U''E^2) + \tfrac{1}{3}\tfrac{k^2}{a^2}\left[4H\Sigma + q\right]
\\
&~~~-\tfrac{1}{3}\left[-U' +4\xi E\right]i\tfrac{k_i}{a}V^i
+Hi\tfrac{k_i}{a}C^i,\end{split}\\
(2\dot{\phi}-\dot{\psi})&=-H(2{\phi}-{\psi})
+\tfrac{1}{4\kappa}\tfrac{a}{4k^2}\left[
-9Hik_jC^j - 3ik_j\dot{C}^j + 2(U' +2\xi E)ik_j V^j\right],\\
\dot{\Sigma}&=-2H\Sigma +\tfrac{1}{2}q
+\tfrac{a}{4k^2}\left[
-9Hik_jC^j - 3ik_j\dot{C}^j + 2(U' +2\xi E)ik_j V^j\right].
\end{align}\label{Eq_first_dynamicfluid}
\end{subequations}

\subsection{Solving the Equations of motion}

Our goal in this section is to solve the Linearized Projected Dirac equations given in Eq.~\eqref{Eq_first_Dir_Proj}, together with the first order Einstein equations listed in Eq.~\eqref{Eq_first_Eins}. We consider the simple potential $U=m E$, for which we found a bouncing background solution in section~\ref{Parityinv}. To make progress we find it also useful to work in a convenient basis $k=\{k_1,0,0\}$, for which the scalar and tensor perturbations of the metric can be written:
\begin{align}
B_i&=
\begin{pmatrix}
0\\
B_2\\
B_3
\end{pmatrix},&
h_{ij} &= \begin{pmatrix}
0&0&0\\
0&h_{22} & h_{23}\\
0& h_{23}& -h_{22}
\end{pmatrix}.\label{BHFORM}
\end{align}
If we were dealing with a model in which only scalar matter were present, then selecting such a basis would be completely without loss of generality. This is not the case for models with spinor content however. As discussed in section~\ref{Sec_Background_FLRW}, the background axial vector $A_{(0)}^I$ picks out a preferred spacelike direction. As a result, the dynamics of the first order perturbations will depend heavily on the orientation of the wave vector $k$ relative to direction picked out by the background. We  therefore separate our this first analysis into two parts: (i) to start with, we analyze `longitudinal' modes for which the wave vector $k$ is aligned with the direction picked out by the axial current, and (ii) we analyze the first order equations for orthogonal modes which lie in the plane perpendicular to the direction picked out by the background axial vector.  

\subsubsection{Longitudinal modes}

We begin by considering perturbative modes for which the wave vector $k$ is aligned with the background axial vector. Given the basis chosen in Eq.~\eqref{BHFORM}, this corresponds to a parity invariant background solution in which $A_{(0)}^2=A_{(0)}^3=0$, and for which $A_{(0)}^1=-E_{(0)}$. From the definitions of the spinor bilinears given in in Eqs.~\eqref{Eq_Bilin_0} and~\eqref{Eq_Bilin_1}, it then follows that for parity invariant backgrounds the following identifications hold:
\begin{align}
\begin{split}
A_{(1)}^1&=-E_{(1)},\\
\widetilde{A}_{(1)}^1&= -\widetilde{E}_{(1)}, \end{split}
&
\begin{split}
V_{(1)}^1&=-A_{(1)}^0,\\
B_{(1)}&=C_{(1)}^1,
\end{split} 
&
\begin{split}
\widetilde{A}_{(1)}^3&=-A_{(1)}^2,\\
V_{(1)}^3 &=C_{(1)}^2,
\end{split}
&
\begin{split}
 \widetilde{A}_{(1)}^2&=A_{(1)}^3,\\
   V_{(1)}^2&=-C_{(1)}^3.
   \end{split}
\end{align}
Making use of these relations allows us to remove much of the degeneracy occurring in the linearized equations of motion. In particular, the $16$ components of the projected Dirac equations given in Eq.~\eqref{Eq_first_Dir_Proj} reduce to the following set of eight:
\begin{subequations}
\begin{align}
\dot{E}_{(1)}&= -3HE_{(1)} +3\dot{\phi}E_{(0)}+ i\tfrac{k_1}{a} A^0_{(1)},\label{AsE}\\
\begin{split}
\dot{\widetilde{E}}_{(1)}&=-3H \widetilde{E}_{(1)} -i\tfrac{k_1}{a}C^1_{(1)} -2(U_{(1)}' + 2\xi E_{(1)})E_{(0)} - 2\psi(U'_{(0)} + 2\xi E_{(0)})E_{(0)},\end{split}\label{EqtwidE}
\\
\dot{A}^0_{(1)}&=- 3H A^0_{(1)}+2U_{(0)}'C^1_{(1)}   +
i\tfrac{k_1}{a} E_{(1)}+[i\tfrac{k_1}{a}\psi - 2i\tfrac{k_1}{a}\phi]E_{(0)},\\
\begin{split}
\dot{C}^1_{(1)}&=-3HC^1_{(1)}-i\tfrac{k^1}{a} \widetilde{E}_{(1)}- 2(U_{(0)}'+2\xi E_{(0)}) A^0_{(1)}
,\end{split}\label{AsC1}\\
\begin{split}
\dot{C}^2_{(1)}&=-3HC^2_{(1)}- 2(U_{(0)}'+2\xi E_{(0)}) C^3_{(1)}
-i\tfrac{k_1}{a}A^{3}_{(1)} -\tfrac{1}{2}\dot{B}_3 E_{(0)}
+(U_{(0)}'E_{(0)}+2\xi E^2_{(0)})B^2,\end{split}\label{ccc2}\\
\begin{split}
\dot{C}^3_{(1)}&=-3HC^3_{(1)}+ 2(U_{(0)}'+2\xi E_{(0)}) C^2_{(1)}
+i\tfrac{k_1}{a}A^{2}_{(1)}
+\tfrac{1}{2}\dot{B}_2E_{(0)}
+(U_{(0)}'E_{(0)}+2\xi E^2_{(0)})B^3,\end{split}\label{ccc3}\\
\dot{A}^2_{(1)}
&= -3HA^2_{(1)}+
i\tfrac{k_1}{a}C^{3}_{(1)},\label{As2}\\
\dot{A}^3_{(1)}
&= -3HA^3_{(1)}-
i\tfrac{k_i}{a}C^{2}_{(1)}.\label{As3}
\end{align}
\end{subequations}
We note rather curiously that for longitudinal modes there is an absence of mixing terms between scalar, vector, and tensor modes exhibited by the projected Dirac equations. To be clear, in the above $8$ equations the `scalar' terms $\phi,~\psi,~ E_{(1)}, ~A_{(1)}^0, ~\widetilde{E}_{(1)}$, and $C_{(1)}^1$ decouple completely from what we will term the components of `vector' modes $B^2,~B^3,~C_{(1)}^2,~C_{(1)}^3, ~A_{(1)}^2,$ and $A_{(1)}^3$.  A similar decoupling also occurs for the Einstein equations, and so for longitudinal modes the SVT decomposition appears to hold, just as would be the case for scalar matter content. The scalar Einstein equations given in Eqs.~\eqref{Eq_first_Eins00} and Eq.~\eqref{Kdata}, become:
\begin{subequations}
\begin{align}
0
&=8\kappa[\tfrac{k_ik^i}{a^2}\phi + 3H\dot{\phi}+ 3H^2\psi]+
(U+  \xi E^2)+\tfrac{i}{2}\tfrac{1}{a}k_1 C^1,  \label{firsteins00}\\%
0&=\widetilde{E}+16\kappa (\dot{\phi} + H\psi),\label{Etwidle}\\
0
&=8\kappa
\left( \ddot{\phi} + (2\dot{H}+ 3H^2)\psi + H(\dot{\psi} + 3\dot{\phi}) \right)-\tfrac{i}{2}\tfrac{k^1}{a}C_1-\xi E^2,\label{einsteiniia}\\
0
&=8\kappa (\phi-\psi) +\tfrac{i}{k^kk_k}a(k_i C^i).\label{einsteiniia2}
\end{align}
\end{subequations}
Similarly, Eqs.~\eqref{Eq_first_einijnewall} yield the following relations for the vector components $B^i$:
\begin{align}
\begin{split}
B^2&=i\tfrac{a}{8\kappa k_1}A_{(1)}^{3},\\
C_{(1)}^{2}&= -4\kappa
(\dot{B}^{2}+2HB^{2}),
\end{split} &
\begin{split}
 B^3&=-i\tfrac{a}{8\kappa k_1}A_{(1)}^{2}%
, \\
 C_{(1)}^{3}&= -4\kappa
(\dot{B}^{3}+2HB^{3}), \end{split}\label{Bs_and_Cs}
\end{align}
together with two equations for the tensor modes $h_{ij}$: 
\begin{align}
\begin{split}
0
 &=4\kappa\Big[
\ddot{h}_{22}+ 3H\dot{h}_{22}+\tfrac{k^kk_k}{a^2} h_{22} 
\Big]- \dot{h}_{23}E_{(0)}, \\
0
 &=4\kappa\Big[\ddot{h}_{23}+ 3H\dot{h}_{23}+\tfrac{k^kk_k}{a^2} h_{23} 
\Big]+ \dot{h}_{22}E_{(0)}. \end{split}\label{tensors1a}
\end{align}
Because the scalars, vectors, and tensors decouple from one another we are able to solve each set of equations independently.  We begin by first considering the scalar dynamics, which at least naively seem to be overconstrained. That is, the 6 
`scalar' terms $\phi,~\psi,~ E_{(1)}, ~A_{(1)}^0, ~\widetilde{E}_{(1)}$, and $C_{(1)}^1$ appear to be governed by eight equations of motion (four projected Dirac equations together with the four `scalar' Einstein equations). It turns out however the not all $8$ equations are independent. By differentiating Eq.~\eqref{firsteins00}, and then making use of Eqs.~\eqref{firsteins00}, \eqref{Etwidle}, \eqref{AsE}, \eqref{einsteiniia}, and \eqref{einsteiniia2}, we can obtain the projected Dirac equation for $C^1$ given in Eq.~\eqref{AsC1}. Similarly, by differentiating Eq.~\eqref{Etwidle} and then making use of Eqs.~Eq.~\eqref{Etwidle}, \eqref{EqtwidE} we can obtain Eq.~\eqref{einsteiniia}. A complete system of equations describing the scalar dynamics is therefore given by the following two Dirac equations:
\begin{subequations}
\begin{align}
\dot{E}_{(1)}&= -3HE_{(1)} +3\dot{\phi}E_{(0)}+ i\tfrac{k_1}{a} A^0_{(1)},\\
\dot{A}^0_{(1)}&=- 3H A^0_{(1)} +
i\tfrac{k_1}{a} [E_{(1)}+16\kappa U_{(0)}'(\phi-\psi) + (\psi - 2\phi)E_{(0)}],
\end{align}
where we have used Eq.~\eqref{einsteiniia2} to remove all instances of $C_{(1)}^1$ from the above equations. The remaining two scalar Einstein equations are similarly given by:
\begin{align}
0
&=8\kappa[\tfrac{1}{2}\tfrac{k_ik^i}{a^2}(\phi+\psi) + 3H\dot{\phi}+ 3H^2\psi]+
(U+  \xi E^2),  \\%
0
&=8\kappa
\left( \ddot{\phi} + (2\dot{H}+ 3H^2)\psi + H(\dot{\psi} + 3\dot{\phi}) \right)+4\kappa \tfrac{k_1k^1}{a^2}(\phi-\psi)-\xi E^2,
\end{align}
\end{subequations}
While these equations are difficult to solve analytically, they are straightforward to solve numerically. We plot the typical behavior of the scalar perturbations below in figures~\ref{fig:PertEA} and~\ref{fig:Pertphipsi}. As can be seen in the figures, the perturbations grow in amplitude towards the time of the bounce, and decay again afterwards. Thus the bounce is stable to longitudinal perturbations. For long wavelength modes, where the $k$-dependence in the equations may be neglected, the spectrum will be unchanged by the bounce, though the amplitude will depend on the details of the background solution and the initial conditions for the perturbations. 
%%%%%%%%%JLL:sentence added below
Notice also that the metric perturbations $\phi$ and $\psi$ are unequal (and in fact they are approximately opposite to each other over large regions of the solution), and thus the solution involves increased anisotropic stress induced by the spinor bilinears in the vicinity of the bounce.
%%%%%%%%%%%end of changes

\begin{figure}[h]
	\begin{minipage}{0.5\textwidth}
	\includegraphics[width=0.9\textwidth]{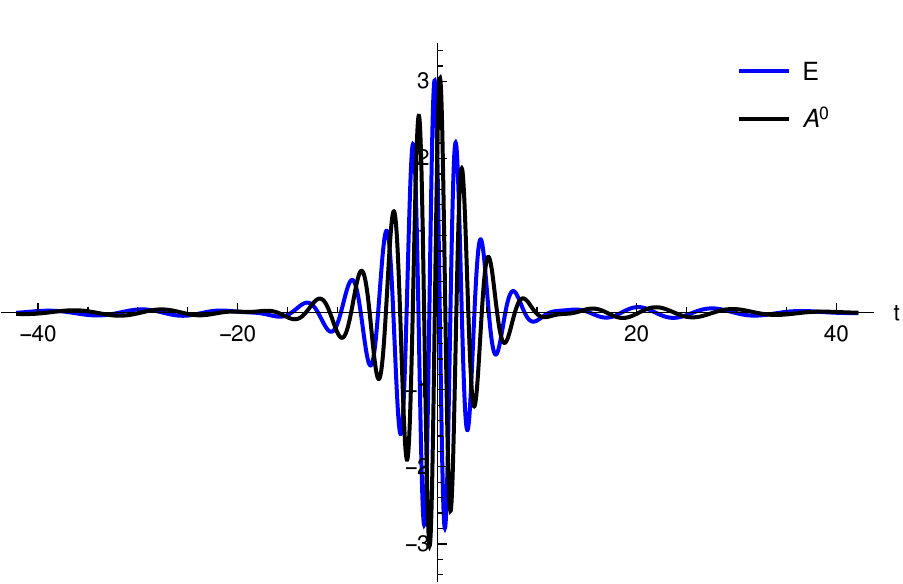}
\end{minipage}%
\begin{minipage}{0.5\textwidth}
	\includegraphics[width=0.9\textwidth]{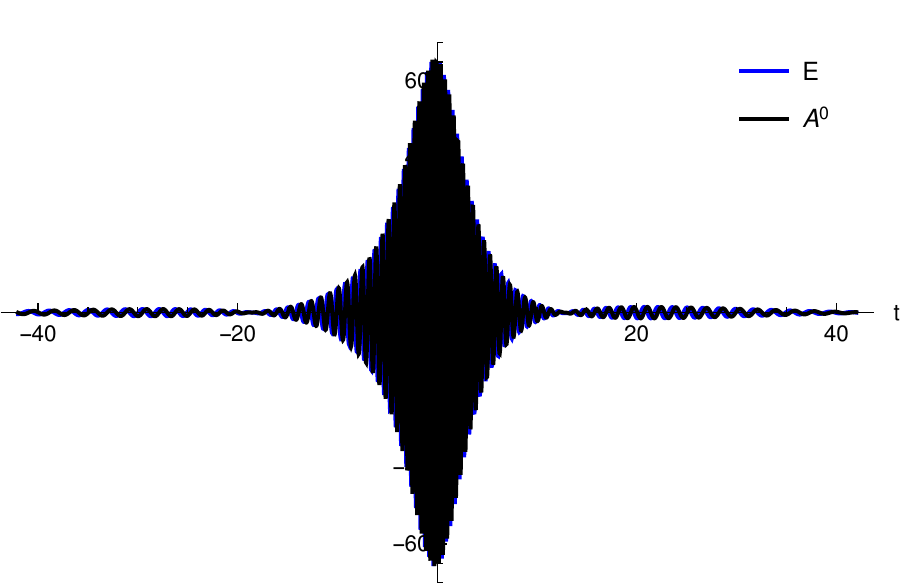}
	\end{minipage}%
	\caption
	{Evolution of the first order spinor bilinears $E_{(1)}$ and $A_{(1)}^1$ for $k_1=2$ (left panel) and $k_1 = 10$ (right panel), for a given choice of initial conditions. We have chosen the potential $U = mE$, with $m=0.1$, $\xi = -0.1$,  $M=0.3$, and $\kappa = 1/4$.  We set the initial conditions at $t=-50$, and choose them such that all Fourier modes are either purely real or purely imaginary, with $-iA_{(1)}^0=E_{(1)}=\phi = \dot{\phi}=\psi=0.1E_{(0)}$.}
	\label{fig:PertEA}
\end{figure}

\begin{figure}[h]
	\begin{minipage}{0.5\textwidth}
	\includegraphics[width=0.9\textwidth]{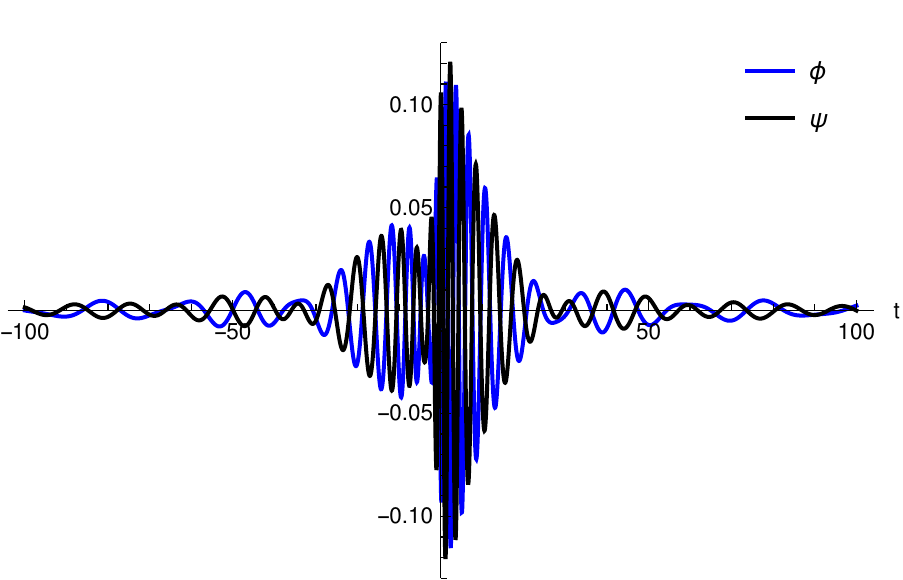}
\end{minipage}%
\begin{minipage}{0.5\textwidth}
	\includegraphics[width=0.9\textwidth]{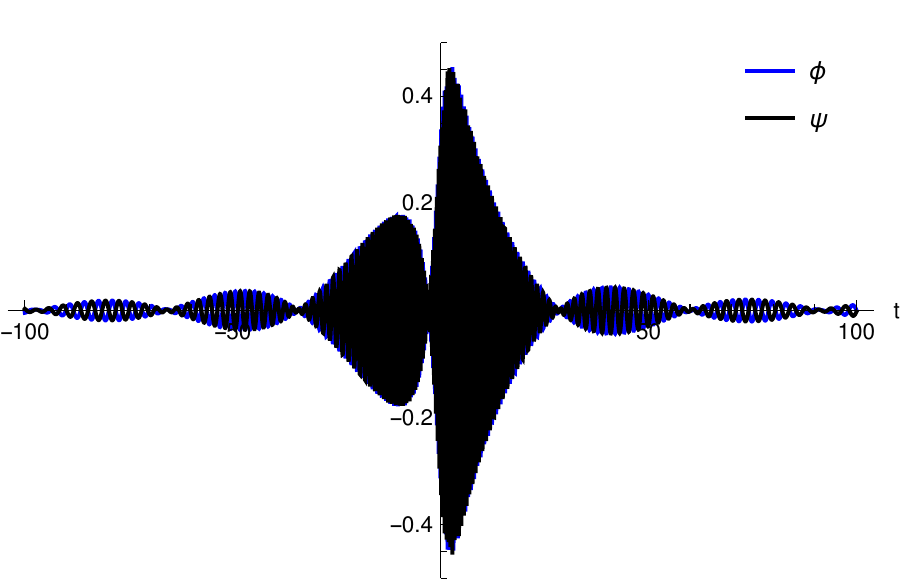}
	\end{minipage}%
	\caption
	{Evolution of the scalar modes $\phi$ and $\psi$ for $k_1=2$ (left panel) and $k_1 = 10$ (right panel), with the same parameter choices as in Fig. \ref{fig:PertEA}.}
	\label{fig:Pertphipsi}
\end{figure}

We next consider the vector perturbations, which once again seem at least naively to be over constrained. That is, the dynamics of the six  modes $C_{(1)}^2,C_{(1)}^3, A_{(1)}^2, A_{(1)}^3, B^2,$ and $B^3$, seem to be governed by $8$ equations of motion (four projected Dirac equations together with the four equations given in~\eqref{Bs_and_Cs}).  Combining Eqs.~\eqref{Bs_and_Cs} together with Eq.~\eqref{As2} and \eqref{As3}, it can be shown after a little manipulation that $C_{(1)}^2 = C_{(1)}^3 = 0$, and therefore also that the $B^i$ scale as $a^{-2}$. The first order axial currents $A^2$ and $A^3$ scale as $a^{-3}$. We therefore find:
\begin{align}
\begin{split}
B^2&=i\tfrac{1}{8\kappa k_1}\tfrac{\widetilde{\alpha}^3}{a^2},\\
A_{(1)}^{2}&= \tfrac{\widetilde{\alpha}^2}{a^3},
\end{split} &
\begin{split}
 B^3&=-i\tfrac{1}{8\kappa k_1}\tfrac{\widetilde{\alpha}^2}{a^2}%
, \\
A_{(1)}^3&= \tfrac{\widetilde{\alpha}^3}{a^3}, \end{split}\label{Bs_and_Cs2}
\end{align}
where the $\widetilde{\alpha}^i$, are $k$-dependent, but time independent. We need to ensure that these solutions are compatible with Eqs.~\eqref{ccc2} and \eqref{ccc3}, which after setting $C^3=C^2=0$, become:
\begin{align}
\begin{split}
0&=8\kappa\tfrac{k_1k^1}{a^2} \widetilde{\alpha}^3+ HE_{(0)}\widetilde{\alpha}^2
+8\kappa\dot{H}\widetilde{\alpha}^3,\end{split}\\
\begin{split}
0&= 8\kappa\tfrac{k_1k^1}{a^2}\widetilde{\alpha}^2
-HE_{(0)}\widetilde{\alpha}^3
+8\kappa\dot{H}\widetilde{\alpha}^2.\end{split}
\end{align}
We see immediately that in order to satisfy these equations, the (k-dependent) constants $\widetilde{\alpha}^2$ and $\widetilde{\alpha}^3$ must be set identically to zero. The `vector' dynamics for longitudinal modes is therefore completely constrained and we have $B^2=B^3=C_{(1)}^2=C_{(1)}^3= A_{(1)}^2=A_{(1)}^3=0$. This situation is reminiscent of the results of Isham and Nelson in~\cite{Isham:1974ci}, who found that solving the full set of Einstein equations for FLRW metrics with spatial curvature required the background axial vector current to be set to zero.

\begin{figure}[h]
	\begin{minipage}{0.5\textwidth}
	\includegraphics[width=0.9\textwidth]{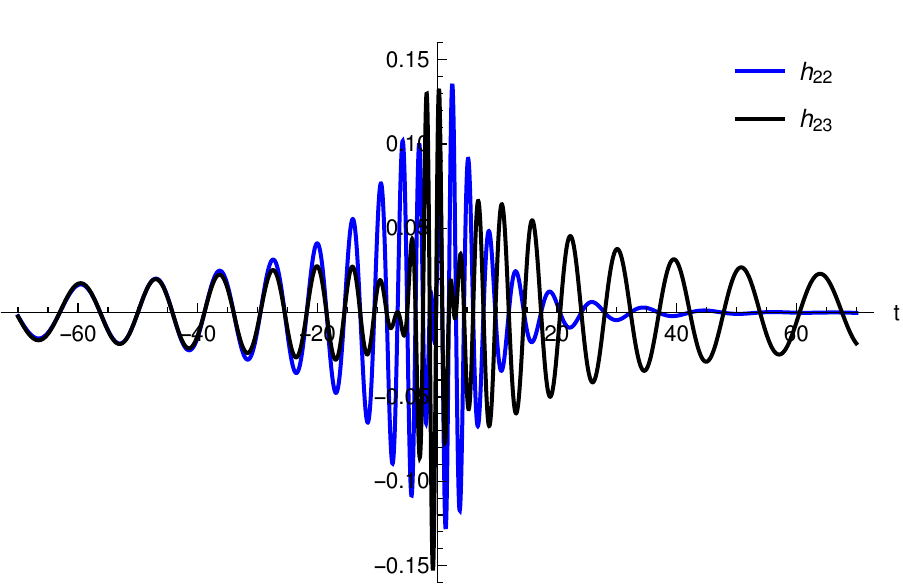}
\end{minipage}%
\begin{minipage}{0.5\textwidth}
	\includegraphics[width=0.9\textwidth]{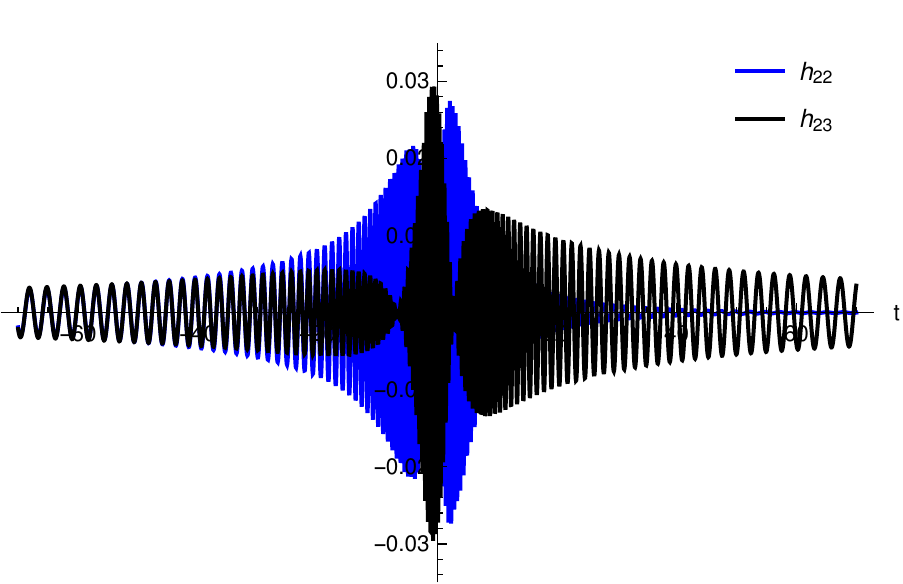}
	\end{minipage}%
	\caption
	{A typical example of the behaviour of tensor modes around the time $t=0$ of the non-singular bounce for $k_1=2$ (left panel) and $k_1 = 10$ (right panel).  We set the initial conditions at $t=-50$, with $h_{22}=\dot{h}_{22}=h_{23}=\dot{h}_{23}=0.01E_{(0)}$.	}
	\label{fig:Perthh}
\end{figure}

The dynamics for the tensor modes is described completely by Eqs.~\eqref{tensors1a}. Solving this pair of equations numerically, we obtain a typical solution, which we display in figure~\ref{fig:Perthh}. Again the solutions grow in amplitude in the approach of the bounce, and decay afterwards, such that the amplitude is typically of comparable magnitude on either side of the bounce. 

\subsubsection{Orthogonal modes}

In this subsection we explore the linearized equations of motion for modes which are perpendicular to the direction picked out by the background axial vector. It is convenient to maintain the basis chosen in Eq.~\eqref{BHFORM} for the first order modes, and so we consider a parity invariant background solution in which $A_{(0)}^1=A_{(0)}^3=0$, and for which $A_{(0)}^2=-E_{(0)}$. It follows from Eqs.~\eqref{Eq_Bilin_0} and~\eqref{Eq_Bilin_1}, that for parity invariant backgrounds the following identifications hold:
\begin{align}
\begin{split}
A_{(1)}^2&=-E_{(1)}, \\
\widetilde{A}_{(1)}^2&= -\widetilde{E}_{(1)},
\end{split} &
\begin{split}
V_{(1)}^2&=-A_{(1)}^0,\\
B_{(1)}&=C_{(1)}^2,
\end{split} &
\begin{split}
\widetilde{A}_{(1)}^3&=A_{(1)}^1,\\
V_{(1)}^3 &=-C_{(1)}^1,
\end{split} &
\begin{split}
 \widetilde{A}_{(1)}^1&=-A_{(1)}^3,\\
  V_{(1)}^1&=C_{(1)}^3.
\end{split} 
\end{align}
As occurred with longitudinal modes we find that the original $16$ components of the projected Dirac equations given in Eq.~\eqref{Eq_first_Dir_Proj} collapse down to the following set of  $8$:
\begin{subequations}
\begin{align}
\dot{E}_{(1)}&= -3HE_{(1)} +3\dot{\phi}E_{(0)}- i\tfrac{k_1}{a} C^3_{(1)} ,\label{projbilE}\\
\begin{split}
\dot{\widetilde{E}}_{(1)}&=-3H \widetilde{E}_{(1)} -i\tfrac{k_1}{a}C^1_{(1)} 
-2(U_{(1)}' + 2\xi E_{(1)})E_{(0)} - 2\psi(U'_{(0)} + 2\xi E_{(0)})E_{(0)},\end{split}\label{coinEt0}\\
\begin{split}
\dot{C}^1_{(1)}&=-3HC^1_{(1)}+ 2(U_{(0)}'+2\xi E_{(0)}) C^3_{(1)}
-i\tfrac{k^1}{a} \widetilde{E}_{(1)}+\tfrac{1}{2}\dot{B}_3E_{(0)},\end{split}\label{c1eqa}\\
\begin{split}
\dot{C}^3_{(1)}&=-3HC^3_{(1)}-i\tfrac{k_1}{a}E_{(1)}- 2(U_{(0)}'+2\xi E_{(0)}) C^1_{(1)}
\\
&~~~-[i\tfrac{k_1}{a}\psi - 2i\tfrac{k_1}{a}\phi]E_{(0)}
+(U_{(0)}'E_{(0)}+2\xi E^2_{(0)})B^3,\end{split}\label{eq_rem4}\\
\dot{A}^0_{(1)}&=- 3H A^0_{(1)}+2U_{(0)}'C_{(1)}^2   -
i\tfrac{k_1}{a} A^1_{(1)}+\tfrac{1}{2}\dot{B}_2E_{(0)},\label{eq_rem1}\\
\dot{A}^1_{(1)}
&= -3HA^1_{(1)}-i\tfrac{k^1}{a}A^0_{(1)},\label{eq_rem2}\\
\dot{A}^3_{(1)}
&= -3HA^3_{(1)}-i
\tfrac{k_1}{a}C^{2}_{(1)},\label{projbilA3}\\
\begin{split}
\dot{C}^2_{(1)}&=-3HC^2_{(1)}-i\tfrac{k_1}{a}A^{3}_{(1)}- 2(U_{(0)}'+2\xi E_{(0)}) A^0_{(1)}
+(U_{(0)}'E_{(0)}+2\xi E^2_{(0)})B^2,\end{split} \label{eq_rem3}
\end{align}
\end{subequations}
We note that unlike was the case for longitudinal modes, no obvious 
decoupling seems to occur between `scalar' and `vector' modes. Instead the set of modes $\{E_{(1)},~\widetilde{E}_{(1)},~C_{(1)}^1,~C_{(1)}^3,~\phi,~\psi,~B^3\}$ seem to decouple from the set $\{A_{(1)}^0,~A_{(1)}^1,~A_{(1)}^3,~C_{(1)}^2,~B^2\}$, at least at the level of the Dirac equations. Let us see if this decomposition continues to hold: The scalar Einstein equations given in Eqs.~\eqref{Eq_first_Eins00} and Eq.~\eqref{Kdata}, become:
\begin{subequations}
\begin{align}
\begin{split}
0
&=8\kappa[\tfrac{k_ik^i}{a^2}\phi + 3H\dot{\phi}+3H^2\psi]+
(U+  \xi E^2)+\tfrac{i}{4}\tfrac{1}{a}[-B^{3}k_{1}E+2k_1 C^1],  
\end{split}\label{coin00}\\
0&=\widetilde{E}+16\kappa (\dot{\phi} + H\psi),\label{cointwidE}\\
0
&=8\kappa
\left( \ddot{\phi} + (2\dot{H}+ 3H^2)\psi + H(\dot{\psi} + 3\dot{\phi}) \right)-\tfrac{i}{2}\tfrac{k^1}{a}C_1-\xi E^2,\label{cooinddot}\\
0 \label{diferenceyo} 
&=8\kappa (\phi-\psi)+\tfrac{i}{4k^kk_k}a(k_1B^{3}E+4k_1 C^1),
\end{align}\label{compthesescal}
\end{subequations}
while the equations of motion given in Eqs.~\eqref{Eq_first_einijnewall} yield the following relations for the vector components $B^i$:
\begin{align}
B^2&=i\tfrac{a}{8\kappa k_1}[A^{3}
+\tfrac{1}{2}h_{23}E], &
B^3&=i\tfrac{a}{8\kappa k_1}[E
-3\phi E-
 \tfrac{1}{2}h_{22}E], \label{vecmodes1}
\end{align}
along with two pairs of equations which relate the tensor modes $h_{ij}$ to the scalar and vector modes:
\begin{subequations}
\begin{align}
 \begin{split}
0
 &= \Big[i\tfrac{1}{2a}C^{2}k_{1}-\tfrac{1}{4}\dot{h}_{23}E\Big]
 +4\kappa 
i\tfrac{1}{2a}(\dot{B}^{2}k_{1}+2HB^{2}k_{1})
,  \\
0
 &= \Big[i\tfrac{1}{2a}C^{3}k_{1}+\tfrac{1}{4}\dot{h}_{22}E\Big]
 +4\kappa 
i\tfrac{1}{2a}(\dot{B}^{3}k_{1}+2HB^{3}k_{1})
 ,\end{split}\label{tenmodes1} \\
\begin{split}
0
 &= \tfrac{1}{2}iB^{3}\tfrac{k_{1}}{a}E
 -4\kappa (\ddot{h}_{22}+ 3H\dot{h}_{22}+\tfrac{k^kk_k}{a^2} h_{22}),\\
0%
 &= \tfrac{1}{2}iB^{2}\tfrac{k_{1}}{a}E+4\kappa(
\ddot{h}_{23}+ 3H\dot{h}_{23}+\tfrac{k^kk_k}{a^2} h_{23} 
). \end{split}\label{compatreq3}
\end{align}
\end{subequations}
Indeed, just as was the case for the projected Dirac equations, no explicit mixing occurs in the Einstein equations between the set 
 $\{E_{(1)},~\widetilde{E}_{(1)},~C_{(1)}^1,~C_{(1)}^3,~\phi,~\psi,~B^3,~h_{22}\}$  and the set $\{A_{(1)}^0,~A_{(1)}^1,~A_{(1)}^3,~C_{(1)}^2,~B^2,~h_{23}\}$. We are free therefore to treat these two sets as behaving  completely independently from one another. Before making use of this decomposition however, we note that a few of the above equations offer immediate analytic solutions. Differentiating Eqs.~\eqref{vecmodes1} with respect to time, and making use of the projected Dirac equations for $E_{(1)}$ and $A_{(1)}^3$ given in Eqs.~\eqref{projbilE} and~\eqref{projbilA3}, we obtain:
\begin{align}
\dot{B}^2&=-2HB^2 -i\tfrac{a}{8\kappa k_1}[i
\tfrac{k_1}{a}C^{2}_{(1)}
-\tfrac{1}{2}\dot{h}_{23}E],&
\dot{B}^3&=-2HB^3-i\tfrac{a}{8\kappa k_1}[ i\tfrac{k_1}{a} C^3_{(1)}
+
 \tfrac{1}{2}\dot{h}_{22}E].\label{vecscaling}
\end{align}
Substituting these equations into Eqs.~\eqref{tenmodes1} we then find  the following relationship between the vector components $C_{(1)}^2$ and $C_{(1)}^3$, and the tensor modes $h_{ij}$:
\begin{align}
i
\tfrac{k_1}{a}C^{2}_{(1)}
 &= \tfrac{1}{2}\dot{h}_{23}E
, &
i\tfrac{k_1}{a} C^3_{(1)}
 &= -\tfrac{1}{2}\dot{h}_{22}E.\label{compagain}
\end{align}
Making use of these relationships for $C_{(1)}^2$ and $C_{(1)}^3$, we see that Eqs.~\eqref{vecscaling} simplifies further, and the vector modes $B^i$ scale as $a^{-2}$. Equations~\eqref{compagain} also allow us to solve Eqs.~\eqref{projbilE}, and~\eqref{projbilA3}, which yield:
\begin{align}
\begin{split}
A^3_{(1)} &= (-\tfrac{1}{2}h_{23} - 2\tfrac{\widetilde{\alpha}^3}{M})
E_{(0)},\\
E_{(1)}&=(3\phi+\tfrac{1}{2}h_{22}-2\tfrac{\widetilde{M}}{M})E_{(0)},\end{split}&
\begin{split}
B^2&=-i\tfrac{1}{4\kappa k_1}[\tfrac{\widetilde{\alpha}^3}{a^2}], \\
B^3&=-i\tfrac{1}{4\kappa k_1}[\tfrac{\widetilde{M}}{a^2}], \end{split}\label{Eqsolsa1}
\end{align}
where the $\widetilde{\alpha}^3$, and $\widetilde{M}$ are (k-dependent) constants with respect to time. These solutions very clearly exhibit mixing between scalar, vector, and tensor modes. 
 
To find the remaining solutions  it will be useful to consider the mode decomposition which occurs between the various perturbative modes for the full set of equations of motion. We begin in particular, by searching for solutions for the set $\{\widetilde{E}_{(1)},~C_{(1)}^1,~C_{(1)}^3,~\phi,~\psi,~h_{22}\}$, which are compatible with the solutions for $\{E_{(1)},~B^3\}$ already given in Eq.~\eqref{Eqsolsa1}. As was the case for longitudinal modes, it   appears at least naively, as though the remaining equations of motion are over constrained, because there are more equations than degrees of freedom. Not every equation is independent however: Differentiating Eq.~\eqref{coin00} with respect to time, and then making use of Eqs.~\eqref{coin00}, \eqref{projbilE}, \eqref{einsteiniia} and \eqref{einsteiniia2} we can derive Eq.~\eqref{c1eqa} for $C_{(1)}^1$. Similarly By differentiating Eq.~\eqref{cointwidE} with respect to time, and then making use of Eqs.~\eqref{cointwidE} and \eqref{coinEt0} we can derive Eq.~\eqref{cooinddot}. We therefore have only to consider the reduced  set  $\{\phi,~\psi,~C_{(1)}^3,~h_{22}\}$, with dynamics specified by the following equations:
\begin{subequations}
\begin{align}
0
&=8\kappa[\tfrac{1}{2}\tfrac{k_ik^i}{a^2}(\phi+\psi) + 3H\dot{\phi}+3H^2\psi-\dot{H}
(3\phi+\tfrac{1}{2}h_{22}-2\tfrac{\widetilde{M}}{M})]-i\tfrac{k_{1}}{a}\tfrac{3}{8}B^{3}E,\label{whittle1}  \\
0
&=
 \ddot{\phi} + (2\dot{H}+ 3H^2)(\psi+3\phi+\tfrac{1}{2}h_{22}-2\tfrac{\widetilde{M}}{M}) + H(\dot{\psi} + 3\dot{\phi}) +\tfrac{1}{2} \tfrac{k_1^2}{a^2}(\phi-\psi)+\tfrac{i}{64\kappa}\tfrac{k^1}{a}B^{3}E,\label{whittle2}\\
0
 &=i\tfrac{k_1}{a} C^3_{(1)}+\tfrac{1}{2}\dot{h}_{22}E,\label{whittle3}\\
0
 &= \tfrac{1}{2}iB^{3}\tfrac{k_{1}}{a}E
 -4\kappa (\ddot{h}_{22}+ 3H\dot{h}_{22}+\tfrac{k^kk_k}{a^2} h_{22}),\label{whittle4}\\
\dot{C}^3_{(1)}&=-3HC^3_{(1)}-i\tfrac{k_1}{a}(\psi +\phi+\tfrac{1}{2}h_{22}-2\tfrac{\widetilde{M}}{M})E_{(0)}+(U_{(0)}'+2\xi E_{(0)})(\tfrac{3}{2}E_{(0)}B^3- 16\kappa i\tfrac{k_1}{a}(\phi-\psi))\label{whittle5},
\end{align}\label{whittle}
\end{subequations}
where we have made use of our solution for $E_{(1)}$ given in Eq.~\eqref{Eqsolsa1} to simplify these equations further. We could similarly have used the solutions given in Eqs.~\eqref{Eqsolsa1} to remove $B^3$ from the equations. Our ultimate goal is to whittle down Eqs.~\eqref{whittle} to a set of four equations in four unknowns. We are unfortunately not able to do so analytically however, and so we must turn to numerics. We find that the above set of equations are indeed consistent, so long as the vector perturbation $B^3$ is set to zero. In particular, we have been able to solve equations~\eqref{whittle1},~\eqref{whittle2},~\eqref{whittle3}, and ~\eqref{whittle4} simultaneously, while making use of Eq.~\eqref{whittle5} in order to set the initial conditions. We then find that Eq.~\eqref{whittle5} remains consistent with the solutions obtained throughout their evolution.  Example solutions for the modes $\{\phi,~\psi,~C_{(1)}^3,~h_{22}\}$ are displayed below in Figs.~\ref{fig:phipsinew}, \ref{fig:A0A1}, \ref{fig:C2C3}, and \ref{fig:H22H23}. All examples we have explored show the same characteristic behavior: the perturbations grow in amplitude towards the bounce, reach a finite maximum value and decay again in a more or less time-symmetric manner after the bounce, though in some cases the amplitude is enhanced by the bounce. Note that in all cases the evolution of scalar, vector and tensor modes is linked, and the presence of scalar modes necessarily induces the presence of gravitational waves. This is one of the main distinguishing features of spinor bounces. 
%%%%%%%%%%%JLL:added sentence below
Also, just as for the longitudinal solutions, the scalar metric perturbations $\phi$ and $\psi$ are unequal, a feature that reveals how near the bounce anisotropic stress plays an increasingly important role. 

We next consider the dynamics of the  four modes $\{A_{(1)}^0,~A_{(1)}^1,~C_{(1)}^2,~h_{23}\}$, which do not include scalar metric perturbations and are completely described by the equations:
\begin{subequations}
\begin{align}
\dot{A}^0_{(1)}&=- 3H A^0_{(1)}+2U_{(0)}'C_{(1)}^2   -
i\tfrac{k_1}{a} A^1_{(1)}-HB_2E_{(0)}\label{eqsub3},\\
\dot{A}^1_{(1)}
&= -3HA^1_{(1)}-i\tfrac{k^1}{a}A^0_{(1)},\label{eqsub1}\\
0
 &=i
\tfrac{k_1}{a}C^{2}_{(1)}- \tfrac{1}{2}\dot{h}_{23}E\label{eqsub4}
,\\
0%
 &= \tfrac{1}{2}iB_{2}\tfrac{k_{1}}{a}E+4\kappa(
\ddot{h}_{23}+ 3H\dot{h}_{23}+\tfrac{k^kk_k}{a^2} h_{23} 
) ,\label{eqsub5}\\
\dot{C}^2_{(1)}&=-3HC^2_{(1)}+i\tfrac{k_1}{a}(\tfrac{1}{2}h_{23} + 2\tfrac{\widetilde{\alpha}^3}{M})
E_{(0)}+(U_{(0)}'+2\xi E_{(0)})(E_{(0)}B^2-2A_{(1)}^0),\label{eqsub2}
\end{align}
\end{subequations}
where we have simplified the above expressions by making use of the  solution for $A_{(1)}^3$ given in Eq.~\eqref{Eqsolsa1}. We could similarly have used the solutions given in Eq.~\eqref{Eqsolsa1} to remove $B^2$. Once again we are not able to continue analytically, however the above five equations are consistent, so long as the vector perturbation $B^2$ is set to zero. In particular, we are able to first solve the equations of motion and constraints~\eqref{eqsub3},~\eqref{eqsub1},~\eqref{eqsub4}, and ~\eqref{eqsub5}, making use of Eq.~\eqref{eqsub2} in order to set the initial conditions. We then find that Eq.~\eqref{eqsub2} remains consistent with the solutions obtained. Example solutions for the modes $\{A_{(1)}^0,~A_{(1)}^1,~C_{(1)}^2,~h_{23}\}$ are displayed below in Figures~\ref{fig:phipsinew}, \ref{fig:A0A1}, \ref{fig:C2C3}, and \ref{fig:H22H23}. These examples are remarkably similar to the solutions obtained for the independent set of modes discussed above. Once again, the bounce is stable and combines scalars, vectors and tensors together. 
\begin{figure}[h]
	\begin{minipage}{0.5\textwidth}
	\includegraphics[width=0.9\textwidth]{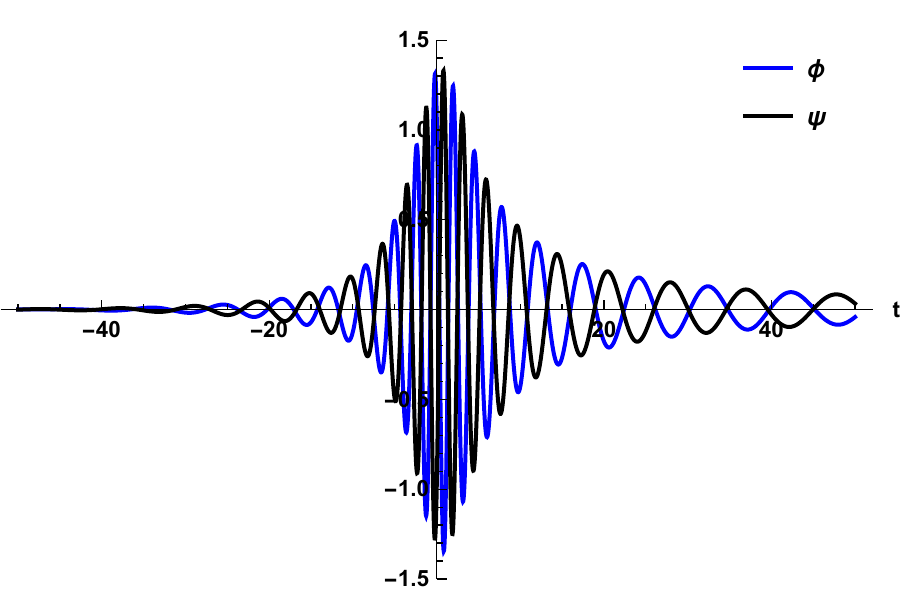}
\end{minipage}%
\begin{minipage}{0.5\textwidth}
	\includegraphics[width=0.9\textwidth]{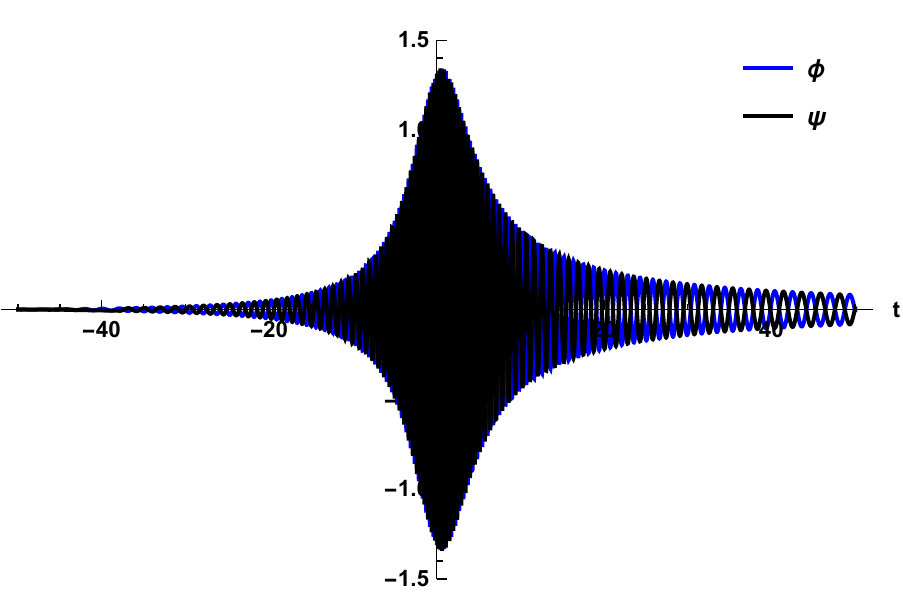}
	\end{minipage}%
	\caption
	{A typical example of the behaviour of the orthogonal scalar modes $\phi$ and $\psi$ around the time $t=0$ of the non-singular bounce for $k_1=2$ (left panel) and $k_1 = 10$ (right panel). 	We set initial conditions at $t=-50$, and again choose initial conditions such that the Fourier modes of the perturbations are either purely real or purely imaginary, with $iA^0_{(1)}=A^1_{(1)}=h_{23}=0.3 E_{(0)}$ and $\phi=-\psi=h_{22}=0.1 E_{(0)}$, while our choice of parameters are given by $m=0.1$, $\xi=-0.1$, and $\kappa = 1/4$.	}
	\label{fig:phipsinew}
\end{figure}

\begin{figure}[h]
	\begin{minipage}{0.5\textwidth}
	\includegraphics[width=0.9\textwidth]{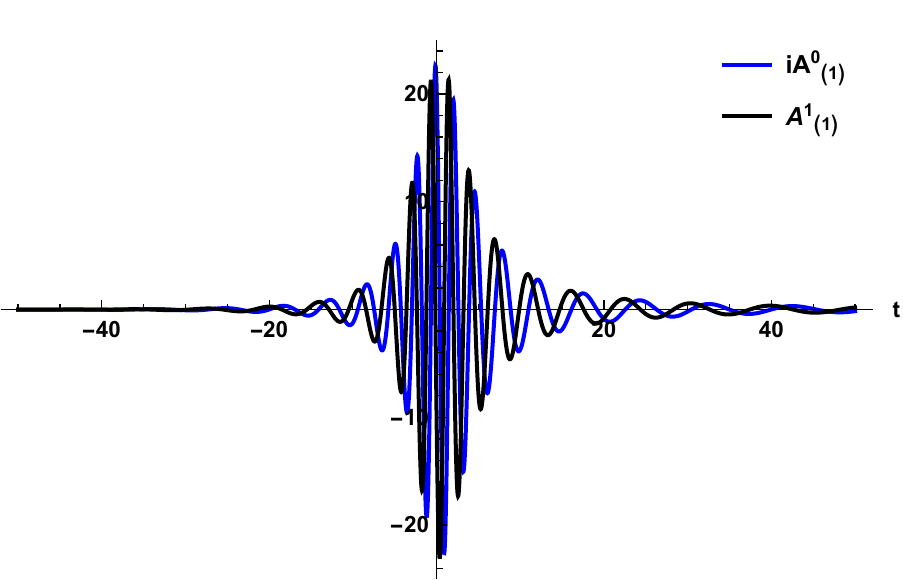}
\end{minipage}%
\begin{minipage}{0.5\textwidth}
	\includegraphics[width=0.9\textwidth]{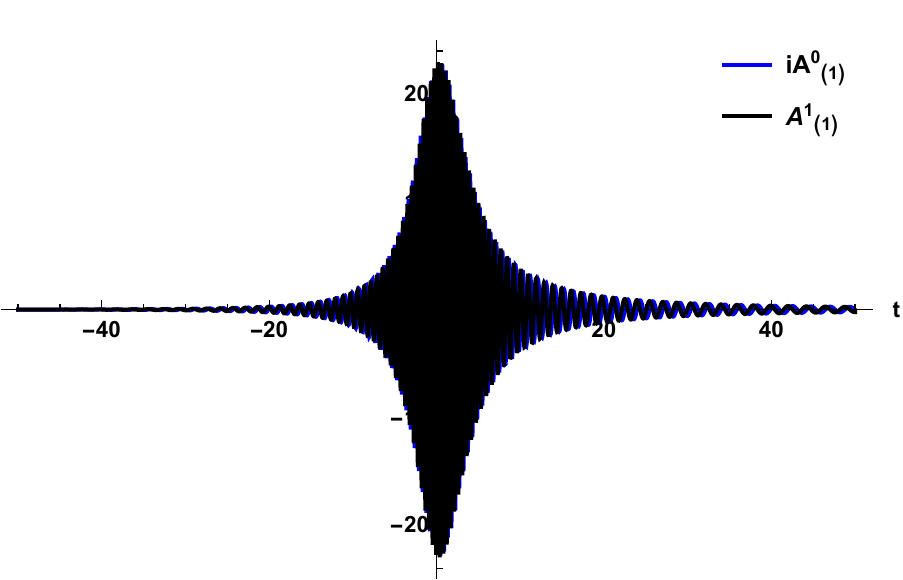}
	\end{minipage}%
	\caption
	{A typical example of the behaviour of the `scalar' modes $i A_{(1)}^0$ and $A_{(1)}^1$ around the time $t=0$ of the non-singular bounce for $k_1=2$ (left panel) and $k_1 = 10$ (right panel), with the same choice of parameters and initial conditions as in Fig. \ref{fig:phipsinew}. }
	\label{fig:A0A1}
\end{figure}

\begin{figure}[h]
	\begin{minipage}{0.5\textwidth}
	\includegraphics[width=0.9\textwidth]{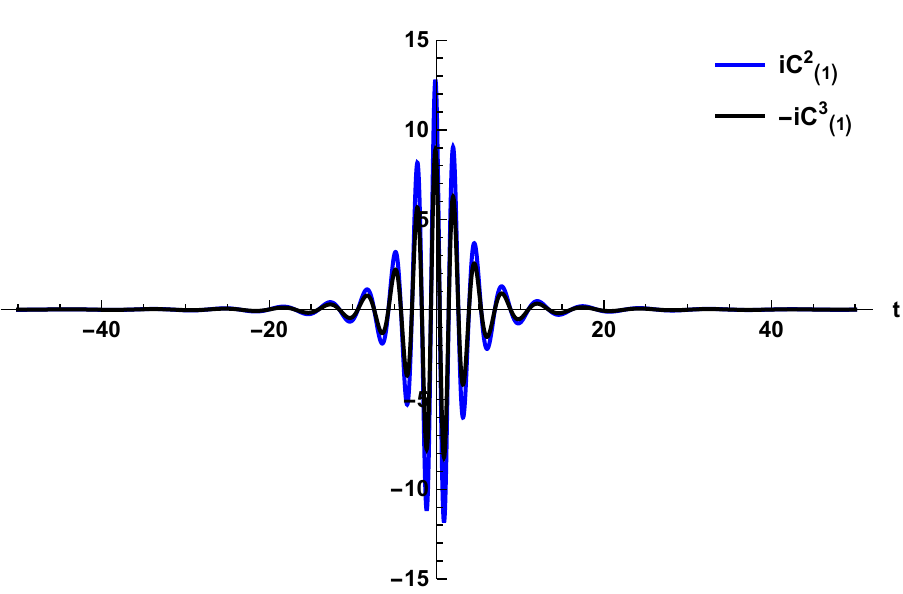}
\end{minipage}%
\begin{minipage}{0.5\textwidth}
	\includegraphics[width=0.9\textwidth]{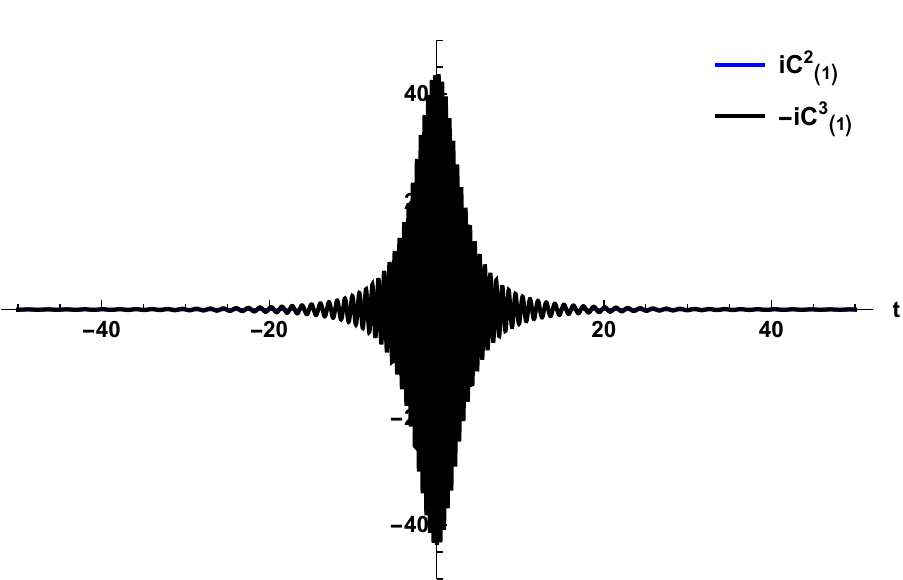}
	\end{minipage}%
	\caption
	{A typical example of the behaviour of the `vector' modes $iC^2_{(1)}$ and $-iC^3_{(1)}$ around the time $t=0$ of the non-singular bounce for $k_1=2$ (left panel) and $k_1 = 10$ (right panel), with the same choice of parameters and initial conditions as in Fig. \ref{fig:phipsinew}. 	}
	\label{fig:C2C3}
\end{figure}

\begin{figure}[h]
	\begin{minipage}{0.5\textwidth}
	\includegraphics[width=0.9\textwidth]{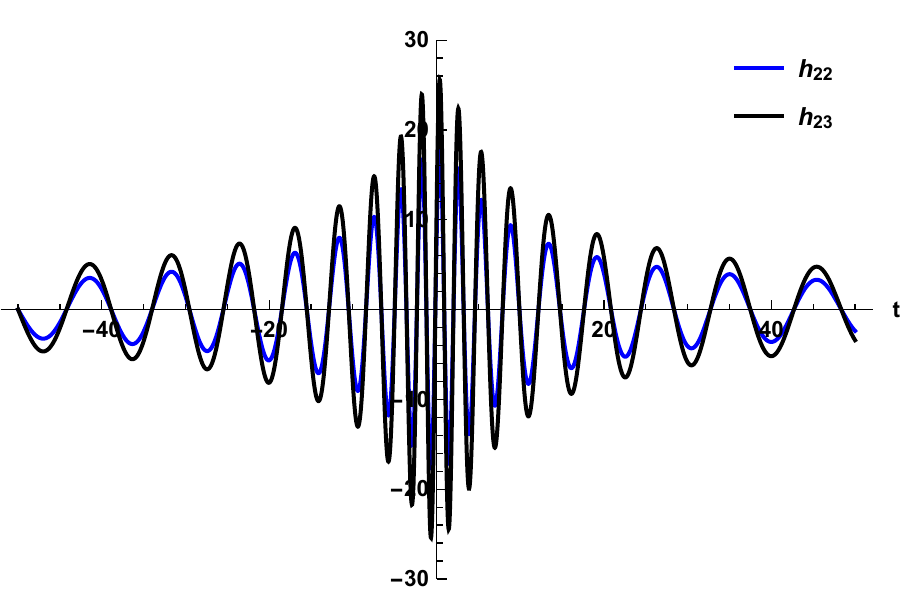}
\end{minipage}%
\begin{minipage}{0.5\textwidth}
	\includegraphics[width=0.9\textwidth]{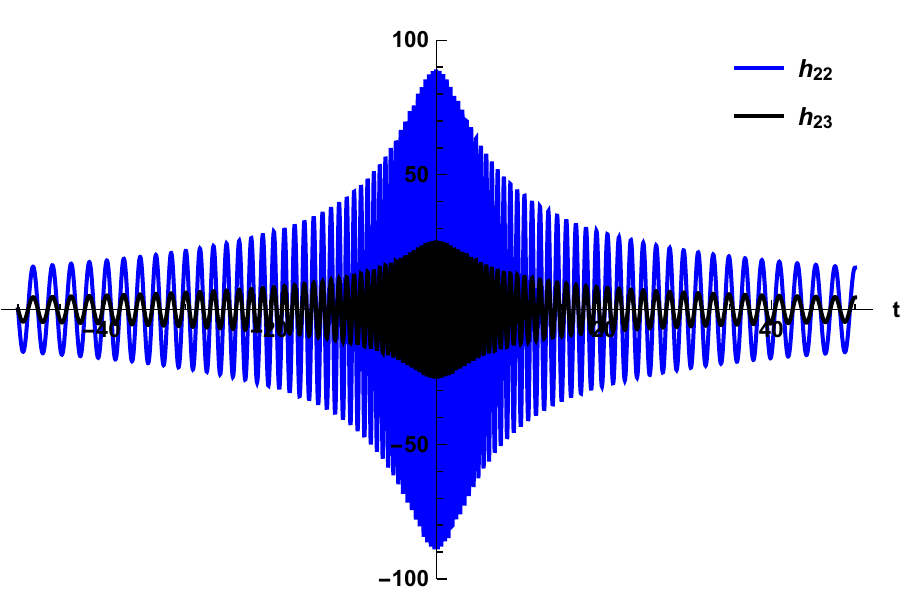}
	\end{minipage}%
	\caption
	{A typical example of the behaviour of the tensor modes $h_{22}$ and $h_{23}$ around the time $t=0$ of the non-singular bounce for $k_1=2$ (left panel) and $k_1 = 10$ (right panel), with the same choice of parameters and initial conditions as in Fig. \ref{fig:phipsinew}. }
	\label{fig:H22H23}
\end{figure}

\section{Discussion}
\label{Sec_Discussion}

Circumventing the big bang is an ambitious goal. In classical general relativity it amounts to finding a loophole in the Penrose-Hawking singularity theorems, which is no easy feat. In particular it is proving notoriously difficult to violate the null energy condition in a convincing way. Spinor cosmologies have the great advantage that they employ a rather minimal generalisation of general relativity which includes adding torsion. This is well motivated from the framework of deriving relativity as a gauge theory, and naturally allows for the coupling of fermions to gravity. It is therefore doubly interesting that such models can also allow for solutions undergoing a cosmological bounce in which the scale factor transitions from a contracting to an expanding phase of the universe. More generally, such models easily allow for a wide range of possible cosmological evolutions, as we have discussed. Nevertheless, let us repeat here that our classicality assumption on spinor bilinears is a rather non-trivial one, which deserves further examination. The results of the present paper motivate a more extensive study where the fermions are treated quantum mechanically, in the spirit of the study of Damour and Spindel \cite{Damour:2011yk}.

%%%%%%%JLL:changed the next two paragraphs
Despite the attractive features of spinor models, to date most treatments of cosmological solutions involved only a discussion of the background. Just a few works looked at perturbations of the spinor fields, but we are not aware of any work taking into account metric perturbations as well. We have done so in this paper, and have discovered several interesting and perhaps unexpected features. For instance, we have found that spinor bounces are stable to linear perturbations, despite the fact that the null energy conditions is violated near the bounce, and this without any particular tuning of the parameters of the models. The stability is ensured by the fact that the torsion terms, which ultimately induce the null energy violating contributions to the stress tensor, only enter algebraically -- thus they do not alter the kinetic terms of the metric or matter perturbations, and do not induce the ghosts that plague many other NEC-violating models.

Furthermore, in cosmological spinor models the standard decomposition into scalar, vector and tensor modes does not in general lead to decoupled equations for these different sets of modes. Rather, the gamma matrices needed to describe Dirac spinors also have the consequence of providing direction-dependent background quantities that spoil the scalar-vector-tensor decomposition usually present at linear order. Far from being a drawback, this feature may well lead to the most interesting consequences for spinor driven bounces: already at linear order, scalar fluctuations generated during a contracting phase will generate gravitational waves. This leads to the interesting prospect that spinor bounces may lead to more pronounced gravitational wave signatures, especially in cosmological models where there is no additional source of primordial gravitational waves at linear order, such as in ekpyrotic models \cite{Boyle:2003km,Lehners:2008vx}. 

Another remarkable consequence of these models is that their cosmological perturbations are direction dependent, even when the background is given by an isotropic FLRW solution. This is again due to the direction dependent gamma matrices required to define the Dirac spinors. A clear goal for future studies will be to determine to what extent these distinguishing features may be observable in cosmological experiments.

\acknowledgments

We would like to thank Abhay Ashtekar, Latham Boyle, Hadi Godazgar, Mingzhe Li, Tom O'brien, Tomas Ortin, Rafael Sorkin, Neil Turok, Edward Wilson-Ewing and Tom Zlosnik for useful discussions and correspondence. T. Q. also wishes to thank the Department of Astrophysics at the University of Science and Technology of China for their hospitality during his visit, during which time this work was finalized. The work of T. Q. is supported in part by the National Natural Science Foundation of China under Grants No. 11405069 and 11653002.

\section{Appendix}
\appendix

\section{Conventions and useful identities}
\label{convens}
In this section we outline the conventions and useful identities that we use in this paper. To facilitate comparison of results we follow closely the conventions used by Magueijo \textit{et al.}~\cite{Magueijo:2012ug}. In particular we work with the flat metric $\eta_{IJ} = diag(-1,1,1,1)$, such that the spacetime metric $g_{\mu\nu} = \eta_{IJ}e^I_\mu e^J_\nu$ is mostly positive. For spacetime indices we use Greek letters, while for internal Lorentz indices we use upper case Roman letters. When summing over only spatial indices we use lower case Roman letters. The co-tetrad is denoted $e^I = e^I_\mu dx^\mu$, while for a general $m-$form $\Lambda$ we define $\Lambda\equiv \tfrac{1}{m!}\Lambda_{a_1...a_m}dx^{a_1}\wedge...\wedge dx^{a_m}$. The determinant of the  co-tetrad  is defined as:
\begin{align}
e = \tfrac{1}{4!}\epsilon_{IJKL}\varepsilon^{\mu\nu\delta\sigma}
e^I_\mu e^J_\nu e^K_\delta e^L_\sigma,\label{Eq_Con_dete}
\end{align} 
where following~\cite{Magueijo:2012ug}, the symbol  $\epsilon_{IJKL}$ is a local $SO(1,3)$ spacetime scalar antisymmetric in all of its indices, and satisfying $\epsilon_{0123}=1$. The symbol $\varepsilon^{\mu\nu\sigma\delta}$ is a spacetime density, which is antisymmetric in all of its indices and satisfies  $\varepsilon^{0123}=1$. Equation~\eqref{Eq_Con_dete}, yields the following useful relation which we make repeated use of throughout the paper:
\begin{align}
\varepsilon^{\mu\nu\delta\sigma}
e^I_\mu e^J_\nu e^K_\delta e^L_\sigma = e\varepsilon^{IJKL} = -e\epsilon^{IJKL}.
\end{align}
We will use the Weyl representation for spinors in this paper, with the following convention for gamma matrices $\gamma^I$:
\begin{align}
\gamma^I &= \begin{pmatrix}
0&\sigma\\
\overline{\sigma} & 0
\end{pmatrix}, & \gamma_5 &= \tfrac{i}{4!}\epsilon_{abcd}\gamma^a\gamma^b\gamma^c\gamma^d =\begin{pmatrix}
-\mathbb{I}_2 & 0\\
0&\mathbb{I}_2
\end{pmatrix} 
\end{align}
where $\sigma = \{1,\sigma^i\}$, $\overline{\sigma} = \{1,-\sigma^i\}$, and where $\sigma^i$ are the Pauli matrices. It is easy to check that these gamma matrices satisfy $\{\gamma^I,\gamma^J\} = -2\eta^{IJ}$. We can use the gamma matrices to construct generators $\Sigma^{ab}$ of the Lorentz group, which  are Krein anti-hermitian and take the form:
\begin{align}
\Sigma^{ab} =-\tfrac{1}{4}[\gamma^a,\gamma^b]\,.
\end{align}
The following identities will be used extensively, and will be indispensable for those wishing to reproduce the results of this paper:
\begin{subequations}
\begin{align}
[[\gamma^a,\gamma^b],\gamma^c] &= 4(\eta^{ac}\gamma^b - \eta^{bc}\gamma^a)\label{identity_gamma_generate}\\
\{[\gamma^a,\gamma^b],\gamma^c\} &= 4\gamma^{[a}\gamma^b\gamma^{c]} = i4\varepsilon^{abcd}\gamma_5\gamma_d\label{identity_gamma_anti-sym}\\
\gamma^0\gamma^a\gamma^0 &= (\gamma^a)^\dagger\label{Gamma_twisted_hermitian}
\end{align}\label{Gamma_identities}
\end{subequations}
The covariant exterior derivative on spinors is given by:
\begin{align}
D\Psi = dx^\mu (\partial_\mu + \omega_\mu)\Psi \equiv dx^\mu (\partial_\mu + \tfrac{1}{2}\omega_{IJ\mu}\Sigma^{IJ})\Psi,
\end{align}
where $\omega$ is the spin connection. The covariant exterior derivative acting on the co-tetrad defines the torsion two form $T^I$:
\begin{align}
T^I\equiv De^I = de^I + \omega^I_{~J}e^J.
\end{align}
As we show in Eq.~\eqref{FullMod_EOM_omega_2}, the spacetime torsion $T^I$ is sourced by the axial and vector currents $A^J$ and $V^J$. It will be useful to decompose the spin connection as follows:
\begin{align}
\omega^{IJ} = \widetilde{\omega}^{IJ}+C^{IJ},
\end{align}
where $\widetilde{\omega}$ is the `torsion free' spin connection obtained by setting $A^I=V^I = 0$. The torsion free spin connection satisfies the equation $T^I=0$, and depends only on the vierbein $e^I$. As a result the torsion can be expressed in terms of the contortion one-form $C^{IJ} = C^{IJ}_{~~~\mu}dx^\mu$: $T^I = C^I_{~J}e^J = C^I_{~JK}e^Ke^J$. We use a `tilde' to indicate when a quantity is `torsion' free. For example the covariant exterior may be denoted
\begin{align}
D = \widetilde{D} +C.
\end{align}
Finally we define the curvature two-form $R^{IJ}$ by:
\begin{align}
R^{IJ} = d\omega^{IJ} + \omega^I_{~K}\omega^{KJ}= \widetilde{R}^{IJ} + \widetilde{D}C^{IJ} + C^I_{~K}C^{KJ}.
\end{align}
The torsion free  Ricci tensor $\widetilde{R}_{\mu\nu}$ and Ricci scalar $\widetilde{R}$ are defined respectively by
\begin{align}
\begin{split}
\widetilde{R}_{\mu\nu} &= e_{I\mu}e_\nu^J\widetilde{R}^{IK}_{~~~JK},\\
\widetilde{R} &= \widetilde{R}^{IJ}_{~~~IJ},
\end{split}
\end{align}
while the torsion free Einstein tensor is defined by
\begin{align}
\widetilde{G}_{\mu\nu} = \widetilde{R}_{\mu\nu}-\tfrac{1}{2}g_{\mu\nu}\widetilde{R}.
\end{align} 

\section{Deriving the equations of motion: A simplified example}

The derivation of the equations of motion is too lengthly to show in complete detail here. Instead we provide the reader with a derivation in full detail for  a simplified model in which $\alpha=\beta=\tau=\lambda=0$, and for which we remove the Holst term in the gravitational part of the action. We also choose $U(\overline{\Psi}\Psi) = 0$. Given these simplifications the full action is written as:
\begin{align}
S &= \kappa\int  \epsilon_{IJKL}e^Ie^J R^{KL}
+ \tfrac{i}{2.3!}\int \epsilon_{IJKL}e^Ie^Je^K(\overline{\Psi}\gamma^L D\Psi - \overline{D\Psi}\gamma^L\Psi).
\end{align}
We consider the vierbein $e^I$, spin connection $\omega^{mn}$, and spinor $\Psi$ as our fundamental fields, and so the  variation of this action is  given by:
\begin{align}
\begin{split}
\delta S &= 2\kappa\int  \epsilon_{IJKL}(\delta e^I) e^J R^{KL}
+ \tfrac{i}{4}\int \epsilon_{IJKL}(\delta e^I)e^Je^K(\overline{\Psi}\gamma^L D\Psi - \overline{D\Psi}\gamma^L\Psi)\\
&
+\kappa\int  \epsilon_{IJKL}e^Ie^J D\delta(\omega^{KL}))
- \tfrac{i}{16.3!}\int \epsilon_{IJKL}e^Ie^Je^K(\delta\omega_{MN})(\overline{\Psi}\{\gamma^L, [\gamma^M,\gamma^N]\}\Psi)\\
&+ \tfrac{i}{2.3!}\int \epsilon_{IJKL}e^Ie^Je^K(\overline{\delta\Psi}\gamma^L D\Psi+\overline{\Psi}\gamma^L D\delta\Psi- \overline{D\Psi}\gamma^L\delta\Psi- \overline{D\delta\Psi}\gamma^L\Psi).\end{split}\label{Eq_simp_var}
\end{align}
The Euler-Lagrange equations are obtained directly from Eq.~\eqref{Eq_simp_var} by setting the variation to zero. Considering first of all the variation of the action with respect to the spin connection we obtain the following equation of motion
\begin{align}
2\kappa  \epsilon_{IJ}^{~~MN}C^I_{~PQ}e^Qe^Pe^J 
= \tfrac{1}{4!} \epsilon_{IJKL}e^Ie^Je^K\varepsilon^{LMNP}A_P,
\end{align}
where we have made use of the identity given in Eq.~\eqref{identity_gamma_anti-sym}. Direct comparison can be made between this result and that which we provided for our full model in Eq.~\eqref{FullMod_EOM_omega_1}. The above equation can be solved to obtain an expression  for the contortion. The approach is to  first multiply through on both sides of the above equation by $e^X$. After some manipulation we then obtain:
\begin{align}
  (
\eta^{X[N}C^{Q|M]}_{~~~Q}
+C^{XNM}) &=\tfrac{1}{16\kappa}\varepsilon^{XNMP}A_P.
\end{align}
Contracting this equation with $\eta_{XM}$ yields $C^{MN}_{~~~M}=0$, and so we immediately find the following algebraic expression for the contortion,
\begin{align}
  C^{XMN} &=\tfrac{1}{16\kappa}\varepsilon^{XMNP}A_P.\label{Eq_Append_Contort}
\end{align}
Direct comparison can be made between this expression for the contortion, and that provided for our full model in Eq.~\eqref{FullMod_EOM_omega_2}. 

We next consider variation of the action with respect to the spinor $\Psi$. After some manipulation, we obtain the following curved space Dirac equation:
\begin{align}
 i \epsilon_{IJKL}\varepsilon^{IJKM}(2\gamma^L \widetilde{D}_M\Psi-\tfrac{1}{8}C_{STM}\{\gamma^L,[\gamma^S,\gamma^T] \Psi)=0
\end{align}
We can write the above equation in completely `torsion free' form, by making use of the expression for the contortion given in Eq.~\eqref{Eq_Append_Contort}. We then find:
\begin{align}
i e^\mu_L\gamma^L \widetilde{D}_\mu\Psi &= \tfrac{i}{8}  C_{MNL}\gamma^L[\gamma^M,\gamma^N]\Psi\nonumber\\
&=  3\pi GA^I\gamma_5\gamma_I\Psi,
\end{align}
where we have again made use of the identity given in Eq.~\eqref{identity_gamma_anti-sym}. If we define $W = \tfrac{3\pi G}{2} A^DA_D$, then we can re-express the Dirac equation as
\begin{align}
i e^\mu_L\gamma^L \widetilde{D}_\mu\Psi 
&= \frac{\delta W}{\delta \overline{\Psi}},\label{Eq_Appen_Dir}
\end{align}
where direct comparison can be made with Eq.~\eqref{FullMod_EOM_omega_4}.

Finally we consider the variation of the action with respect to the vierbein, which yields the following Einstein equation:
\begin{align}
 \kappa  \epsilon_{IJKL} e^Je^Me^N R^{KL}_{~~~MN}
&=- \tfrac{i}{4} \epsilon_{IJKL}e^Je^Ke^M(\overline{\Psi}\gamma^L D_M\Psi - \overline{D_M\Psi}\gamma^L\Psi).
\end{align}
We use the expression for the contortion given in  Eq~\eqref{Eq_Append_Contort} to re-write this equation in torsion free form:
\begin{align}
\begin{split}
 \kappa  \epsilon_{IJKL} e^Je^Me^N \widetilde{R}^{KL}_{~~~MN}
&=- \tfrac{i}{4} \epsilon_{IJKL}e^Je^Ke^M(\overline{\Psi}\gamma^L \widetilde{D}_M\Psi - \overline{\widetilde{D}_M\Psi}\gamma^L\Psi)\\
&~~~+ \tfrac{1}{4.16\kappa} \epsilon_{IJKL}e^Je^Ke^M(\delta_M^LA^PA_P-A^LA_M)\\
&~~~-\tfrac{1}{8}  \epsilon_{IJKL} e^Je^Me^N \varepsilon^{KL~P}_{~~~N}\widetilde{D}_MA_P\\
&~~~-\tfrac{1}{8.16\kappa}  \epsilon_{IJKL} e^Je^Me^N (
 \varepsilon_{PM}^{~~~KQ}A_Q\varepsilon^{PL}_{~~~NS}A^S).\end{split}
\end{align}
The Einstein equations appear very unfamiliar when written in this `first-order' form. To re-express them in second order form we first multiply though on both sides by $e^X$. After some some manipulation we obtain:
\begin{align}
 4\kappa  \widetilde{G}_{\mu\nu}
=\tfrac{i}{2}g_{\mu\nu} \widetilde{X} - \tfrac{3}{4.16\kappa}g_{\mu\nu}A_IA^I  -\tfrac{i}{2}e_{L\mu } \widetilde{X}_{\nu}^L- \tfrac{1}{4}e_{L\mu}e_{M\nu}\varepsilon^{LMNP}\widetilde{D}_NA_P.
\end{align}
where $X^L_\mu = (\overline{\Psi}\gamma^L D_\mu\Psi - \overline{D_\mu\Psi}\gamma^L\Psi)$, and where we have used the vierbein to switch to spacetime indices. Notice that the above equation appears a little strange in that the last two terms on the right hand side do not appear to be symmetric in their indices. Let us however consider the anti-symmetric part of the second term on its own:
\begin{align}
-\tfrac{i}{2}e_{L[\mu } \widetilde{X}_{\nu]}^L&=-\tfrac{i}{4}e_{L\mu } e_{M\nu}(\overline{\Psi}(\gamma^L\eta^{MN}-\gamma^M\eta^{LN}) \widetilde{D}_N\Psi - \overline{\widetilde{D}_N\Psi}(\gamma^L\eta^{MN}-\gamma^M\eta^{LN})\Psi)\nonumber\\
&=-\tfrac{i}{16}e_{L\mu } e_{M\nu}(\overline{\Psi}[[\gamma^M,\gamma^L],\gamma^N]  \widetilde{D}_N\Psi - \overline{\widetilde{D}_N\Psi}[[\gamma^M,\gamma^L],\gamma^N]\Psi)\nonumber\\
&=\tfrac{i}{16}e_{L\mu } e_{M\nu}(\overline{\Psi}\{[\gamma^M,\gamma^L],\gamma^N\}  \widetilde{D}_N\Psi + \overline{\widetilde{D}_N\Psi}\{[\gamma^M,\gamma^L],\gamma^N\}\Psi)\nonumber\\
&~~~-\tfrac{i}{8}e_{L\mu } e_{M\nu}(\overline{\Psi}[\gamma^M,\gamma^L]\gamma^N  \widetilde{D}_N\Psi + \overline{\widetilde{D}_N\Psi}\gamma^N[\gamma^M,\gamma^L]\Psi)\nonumber\\
&=\tfrac{1}{4}e_{L\mu } e_{M\nu}\varepsilon^{LMNP}\widetilde{D}_N A_P-\tfrac{i}{8}e_{L\mu } e_{M\nu}(\overline{\Psi}[\gamma^M,\gamma^L]\gamma^N  \widetilde{D}_N\Psi + \overline{\widetilde{D}_N\Psi}\gamma^N[\gamma^M,\gamma^L]\Psi),
\end{align}
where for the second equality we have made use of the identity given in Eq.~\eqref{identity_gamma_generate}. We therefore express the torsion free Einstein equation in its final form as:
\begin{align}
\begin{split}
 4\kappa  \widetilde{G}_{\mu\nu}
&=\tfrac{i}{2}g_{\mu\nu} \widetilde{X}-\tfrac{i}{2}e_{L(\mu } \widetilde{X}_{\nu)}^L - g_{\mu\nu}\tfrac{3\pi G}{2}A_IA^I  \\
&~~~-\tfrac{i}{8}e_{L\mu } e_{M\nu}(\overline{\Psi}[\gamma^M,\gamma^L]\gamma^N  \widetilde{D}_N\Psi + \overline{\widetilde{D}_N\Psi}\gamma^N[\gamma^M,\gamma^L]\Psi).
\end{split}
\end{align}
Direct comparison can be made between this equation and Eq.~\eqref{FullMod_EOM_omega_7}. Once again the second line on the right hand side does not appear symmetric in its indices. Notice however that this term is identically zero on shell, which can be seen by making use of the Dirac equation given in Eq.~\eqref{Eq_Appen_Dir}.
\label{Sec_SimpleDer}
\section{Deriving the form of the Contortion tensor $C^{IJK}$}

In section~\ref{Sec_SimpleDer} we derived the equations of motion for a simplified model with no Holst term, and for which $\alpha=\beta=\tau=\lambda = 0$. While deriving the equations of motion for the full model is significantly more computationally intensive, it is for the most part no more technically difficult. Perhaps the one exception is in deriving Eq.~\eqref{FullMod_EOM_omega_2} from Eq.~\eqref{FullMod_EOM_omega_1}. In this section we therefore fill in the details of that calculation. We begin by first contracting both sides of Eq.~\eqref{FullMod_EOM_omega_1} with $(\varepsilon^{STMN} -2\gamma \eta^{S[M}\eta^{N]T})e^X$, which after some manipulation yields:
\begin{align}
\begin{split}
-8\kappa\gamma\left(\frac{1+\gamma^2}{\gamma^2}\right)\left[C^{X[TS]} + \eta^{X[T}C^{Q|S]}_{~~~~Q}\right]&= \eta^{X[S}\delta^{T]}_A(\gamma Q^A +(A^A +\overline{Q}^A))\\
&+\tfrac{1}{2}\varepsilon^{QXST}(Q_Q - \gamma(A_Q + \overline{Q}_Q)).\end{split}
\end{align}
The first term on the left hand side of this equation has anti-symmetrised brackets around two of its indices that we wish to remove. In order to do so we sum together instances of Eq.~\eqref{FullMod_EOM_omega_2} with various permutations of its indices. From this procedure we obtain:
\begin{align}
\begin{split}
8\kappa\gamma\left(\frac{1+\gamma^2}{\gamma^2}\right)\left[C^{TXS} + 2\eta^{S[X}C^{Q|T]}_{~~~~Q}\right]&=\eta^{S[X}\delta^{T]}_A(\gamma Q^A +(A^A +\overline{Q}^A))\\
&+\tfrac{1}{2}\varepsilon^{QXST}(Q_Q - \gamma(A_Q + \overline{Q}_Q)).\end{split}
\end{align}
Next, in order to remove the second term on the left hand side of this equation, notice that we can obtain an expression for $C^{TX}_{~~~T}$ on its own by contracting Eq.~\eqref{FullMod_EOM_omega_3} with $\eta_{ST}$:
\begin{align}
C^{TX}_{~~~~T}&=\frac{3\gamma}{16\kappa (1+\gamma^2)}\left[\gamma Q^X + (A^X - \overline{Q}^X)\right].
\end{align}
Substituting this expression for $C^{TX}_{~~~T}$ back into Eq.~\eqref{FullMod_EOM_omega_3} finally yields an algebraic equation for the contortion which we provide in Eq.~\eqref{FullMod_EOM_omega_2}.

\section{Selecting a gauge}

In order to simplify the linearly perturbed equations of motion we choose to work in a fixed gauge. As vierbein gauge fixing is not often discussed in the literature we will provide details here. We begin by determining how our metric and vierbein perturbations transform under the symmetries of our model. The line element in conformal time is given by
\begin{align}
ds^2 = a(\tau)^2[-(1+2\psi)d\tau^2 + 2B_id\tau dx^i + (\delta_{ij} + h_{ij})dx^idx^j]\,,
\end{align}
where $\psi,\phi, B_i,$ and $h_{ij}$ are perturbations which depend a priori on all spacetime coordinates. We can use the  invariance of the line element under coordinate transformations to determine how each of these perturbations transforms under a coordinate transformation:
\begin{align}
ds^2 = g_{\mu\nu}dx^\mu dx^\nu &=
 \widetilde{g}_{\rho\sigma}d\widetilde{x}^\rho d\widetilde{x}^\sigma\nonumber\\
&= \widetilde{g}_{\rho\sigma}\tfrac{d\widetilde{x}^\rho}{dx^\mu}\tfrac{d\widetilde{x}^\sigma}{dx^\nu}dx^\mu dx^\nu\nonumber\\
\longrightarrow g_{\mu\nu}&= \widetilde{g}_{\rho\sigma}\tfrac{d\widetilde{x}^\rho}{dx^\mu}\tfrac{d\widetilde{x}^\sigma}{dx^\nu}\label{Eq_gagfix_metT}
\end{align} 
If we consider a coordinate transformation of the form $x^\mu\rightarrow \widetilde{x}^\mu = x^\mu + \xi^\mu$, where $\xi^\mu$ is small so that it can be treated as a perturbation, Eq~\eqref{Eq_gagfix_metT} tells us that the metric perturbations must transform as:
\begin{subequations}
\begin{align}
\psi\rightarrow \widetilde{\psi} &= \psi - \mathcal{H}\xi^0 - \dot{\xi}^0\\
B_i\rightarrow \widetilde{B}_i &= B_i + \partial_i\xi^0 - \delta_{ij}\dot{\xi}^j\\
h_{ij}\rightarrow \widetilde{h}_{ij} & = h_{ij} - 2\xi^0\mathcal{H}\delta_{ij}- \delta_{kj}\partial_i \xi^k - \delta_{ki}\partial_j \xi^k
\end{align}
\label{Eq_gagfix_metT_List}
\end{subequations}
Following a similar procedure we can also work out how the perturbed vierbein must transform under coordinate transformations. We express the components of the perturbed vierbein $e_\mu^a$ in conformal time as:
\begin{align}
e^0_0  = a(\tau)(1 + \psi),\hspace{1cm} e^0_i = a(\tau)C_i,\hspace{1cm}e^i_0 = a(\tau)E^i,\hspace{1cm}e^i_j= a(\tau)(\delta^i_j + k^i_j).
\end{align}
Note that the $C_i$ and $E^i$ that are used in this section denote spacetime fluctuations, and that they are unrelated to the spinor bilinears used in the rest of the paper.
We then use the invariance of the tetrad $e^a$ under coordinate transformations to determine how the vierbein must transform:
\begin{align}
e^a=e^a_\mu dx^\mu &= \widetilde{e}^a_\nu d\widetilde{x}^\nu\nonumber\\
&=\widetilde{e}^a_\nu \tfrac{d\widetilde{x}^\nu}{dx^\mu}dx^\mu\nonumber\\
\rightarrow e^a_\mu &= \widetilde{e}^a_\nu \tfrac{d\widetilde{x}^\nu}{dx^\mu}\label{Eq_gagfix_vierT}
\end{align}
Once again looking at coordinate transformations of the form $x^\mu\rightarrow \widetilde{x}^\mu = x^\mu + \xi^\mu$ where $\xi^\mu$ is small so that we can treat it like a perturbation, Eq.~\eqref{Eq_gagfix_vierT} tells us that the vierbein perturbations transform as
\begin{subequations}
\begin{align}
\psi\rightarrow \widetilde{\psi} &= \psi - \mathcal{H}\xi^0 - \dot{\xi}^0\,, \\
E^i\rightarrow \widetilde{E}^i &= E^i - \dot{\xi}^i\,, \\
C_i\rightarrow \widetilde{C}_i &= C_i - \partial_i\xi^0\,, \\
k^i_j\rightarrow \widetilde{k}^i_j &= k^i_j - \partial_j\xi^i - \delta^i_j \mathcal{H} \xi^0\,.
\end{align}\label{Eq_gagfix_vierT_List}
\end{subequations}
Notice that the vierbein and metric perturbations are not independent, but are linked via the definition $g_{\mu\nu}= e^a_\mu e^b_\nu n_{ab}$. We therefore find:
\begin{subequations}
\begin{align}
B_i &= -C_i + \eta_{ij}E^j\,, \\
h_{ij} &= \sum_k(\delta^k_ik^k_j + \delta^k_j k^k_i)\,.
\end{align}
\label{Eq_gagfix_viermet}
\end{subequations}
It is easy to check that the gauge transformations given in Eq.~\eqref{Eq_gagfix_metT_List} can be reproduced by making use of Eqs.~\eqref{Eq_gagfix_vierT_List} together with Eqs.~\eqref{Eq_gagfix_viermet}.

As well as a coordinate index, the vierbeine also have a Lorentz index. Under an infinitesimal Lorentz transformation, the vierbeine transform as $e_\mu^a \rightarrow \widehat{e}_\mu^a=(\delta^a_b + \Lambda^a_{\bullet b}) e_\mu^b$. As short calculation shows that vierbein perturbations therefore transform as
\begin{subequations}
\begin{align}
\psi\rightarrow \widehat{\psi} &= \psi,\\
E^i\rightarrow \widehat{E}^i &= E^i +\Lambda^i_{~ 0},\\
C_i\rightarrow \widehat{C}_i &= C_i + \Lambda^0_{~ i},\\
k^i_j\rightarrow \widehat{k}^i_j &= k^i_j + \Lambda^i_{~ j}.
\end{align}
\end{subequations}
From Eqs.~\eqref{Eq_gagfix_viermet} it follows that the metric perturbations $B_i$ and $h_{ij}$ remain invariant under local infinitesimal Lorentz transformations, as expected.

Let us now eliminate the redundant components that arise because we are analyzing a gauge theory. We start by fixing a local Lorentz gauge. The vierbein perturbations account for $16$ degrees of freedom, which is $6$ more than for the metric. The Lorentz group is 6 dimensional, and so the choice we take is to fix the extra degrees of freedom in the vierbein (i.e. we leave the spinor degrees of freedom completely un-fixed). We set:
\begin{align}
C_i &= - \eta_{ij}E^j,\\
k_{ij} &= k_{ji}, 
\end{align} 
which from Eq.~\eqref{Eq_gagfix_viermet} implies $B_i = -2C_i = 2E_i$. Having used up all of our Lorentz freedom to gauge fix the vierbein, we next want to use coordinate freedom to do the same for the metric. Before doing so however it will be useful to first perform a so called scalar, vector, tensor (SVT) decomposition of the metric. This means that we decompose the vector perturbation $B_i$ into its scalar and vector components~\cite{Stewart:1990fm}:
\begin{align}
B_i = \underbrace{\partial_i B}_\text{scalar} + \underbrace{\widehat{B}_i}_\text{vector}
\end{align}
where $\partial^i \widehat{B}_i$ = 0. We similarly decompose $h_{ij}$ into its constituent parts:
\begin{align}
h_{ij} = \underbrace{2\delta_{ij}\phi + 2\partial_{\langle i}\partial_{j\rangle}\chi}_\text{scalar} + \underbrace{2\partial_{(i}\widehat{F}_{j)}}_\text{vector} + \underbrace{2\widehat{h}_{ij}}_{tensor},
\end{align}
where $\partial_{\langle i}\partial_{j\rangle}\chi = (\partial_i\partial_j - \tfrac{1}{3}\delta_{ij}\partial^k\partial_k)\chi$, and hatted quantities are divergence-less $\partial^i \widehat{F}_i = \partial^i\widehat{h}_{ij}=0$ and traceless $\widehat{h}^i_i = 0$. Likewise $\xi^i = \widehat{\xi}^i + \xi^{,i}.$ We therefore see that the metric perturbations decompose into four scalar degrees of freedom $\phi,\psi,B,\chi$, four vector degrees of freedom $\widehat{B}_i,\widehat{F}_i$, and two tensor degrees of freedom $\widehat{h}_{ij}$. Under this decomposition the gauge transformations given in Eq.~\eqref{Eq_gagfix_metT_List} are re-expressed as:
\begin{align}
\begin{split}
\psi&\rightarrow \psi - \mathcal{H}\xi^0 - \dot{\xi}^0,\\
\phi&\rightarrow  \phi -\xi^0\mathcal{H}-\tfrac{1}{3}\partial_k\partial^k\xi,\\
\widehat{B}_i&\rightarrow \widehat{B}_i  - \partial_0{\widehat{\xi}}_i,
\end{split}
&
\begin{split}
\chi &\rightarrow \chi -\xi,\\
B&\rightarrow B+ \xi^0-\dot{\xi},\\
\widehat{F}_i&\rightarrow \widehat{F}_i - \widehat{\xi}_i,
\end{split}
\end{align}
while $\widehat{h}_{ij}$ is gauge invariant. It is immediately clear that by choosing $\xi,\xi^0$ and $\widehat{\xi}_i$ appropriately we have the freedom to set two scalar and two vector degrees of freedom to zero. We choose Newtonian gauge: $B = \chi = \widehat{F}_i=0$. The final line element is given in Eq.~\eqref{lineelement}, where in an effort to reduce clutter we have dropped the `hats' on the vector and tensor perturbations.

\bibliographystyle{utphys}
\bibliography{bibli}

\providecommand{\href}[2]{#2}\begingroup\raggedright\begin{thebibliography}{10}

\bibitem{Vilenkin:1983xq}
A.~Vilenkin, ``{The Birth of Inflationary Universes},''
\href{http://dx.doi.org/10.1103/PhysRevD.27.2848}{{\em Phys. Rev.} {\bfseries
  D27} (1983) 2848}.
%%CITATION = PHRVA,D27,2848;%%.

\bibitem{Hartle:1983ai}
J.~B. Hartle and S.~W. Hawking, ``{Wave Function of the Universe},''
\href{http://dx.doi.org/10.1103/PhysRevD.28.2960}{{\em Phys. Rev.} {\bfseries
  D28} (1983) 2960--2975}.
%%CITATION = PHRVA,D28,2960;%%.

\bibitem{Feldbrugge:2017kzv}
J.~Feldbrugge, J.-L. Lehners, and N.~Turok, ``{Lorentzian Quantum Cosmology},''
  \href{http://dx.doi.org/10.1103/PhysRevD.95.103508}{{\em Phys. Rev.}
  {\bfseries D95} no.~10, (2017) 103508},
\href{http://arxiv.org/abs/1703.02076}{{\ttfamily arXiv:1703.02076 [hep-th]}}.
%%CITATION = ARXIV:1703.02076;%%.

\bibitem{Feldbrugge:2017fcc}
J.~Feldbrugge, J.-L. Lehners, and N.~Turok, ``{No smooth beginning for
  spacetime},''
\href{http://arxiv.org/abs/1705.00192}{{\ttfamily arXiv:1705.00192 [hep-th]}}.
%%CITATION = ARXIV:1705.00192;%%.

\bibitem{DiazDorronsoro:2017hti}
J.~Diaz~Dorronsoro, J.~J. Halliwell, J.~B. Hartle, T.~Hertog, and O.~Janssen,
  ``{Real no-boundary wave function in Lorentzian quantum cosmology},''
  \href{http://dx.doi.org/10.1103/PhysRevD.96.043505}{{\em Phys. Rev.}
  {\bfseries D96} no.~4, (2017) 043505},
\href{http://arxiv.org/abs/1705.05340}{{\ttfamily arXiv:1705.05340 [gr-qc]}}.
%%CITATION = ARXIV:1705.05340;%%.

\bibitem{Feldbrugge:2017mbc}
J.~Feldbrugge, J.-L. Lehners, and N.~Turok, ``{No Rescue for the No Boundary
  Proposal},''
\href{http://arxiv.org/abs/1708.05104}{{\ttfamily arXiv:1708.05104 [hep-th]}}.
%%CITATION = ARXIV:1708.05104;%%.

\bibitem{Gielen:2015uaa}
S.~Gielen and N.~Turok, ``{Perfect Quantum Cosmological Bounce},''
  \href{http://dx.doi.org/10.1103/PhysRevLett.117.021301}{{\em Phys. Rev.
  Lett.} {\bfseries 117} no.~2, (2016) 021301},
\href{http://arxiv.org/abs/1510.00699}{{\ttfamily arXiv:1510.00699 [hep-th]}}.
%%CITATION = ARXIV:1510.00699;%%.

\bibitem{Chen:2016ask}
P.~Chen, Y.-C. Hu, and D.-h. Yeom, ``{Fuzzy Euclidean wormholes in de Sitter
  space},''
\href{http://arxiv.org/abs/1611.08468}{{\ttfamily arXiv:1611.08468 [gr-qc]}}.
%%CITATION = ARXIV:1611.08468;%%.

\bibitem{Bramberger:2017cgf}
S.~F. Bramberger, T.~Hertog, J.-L. Lehners, and Y.~Vreys, ``{Quantum
  Transitions Through Cosmological Singularities},''
\href{http://arxiv.org/abs/1701.05399}{{\ttfamily arXiv:1701.05399 [hep-th]}}.
%%CITATION = ARXIV:1701.05399;%%.

\bibitem{Cai:2016thi}
Y.~Cai, Y.~Wan, H.-G. Li, T.~Qiu, and Y.-S. Piao, ``{The Effective Field Theory
  of nonsingular cosmology},''
  \href{http://dx.doi.org/10.1007/JHEP01(2017)090}{{\em JHEP} {\bfseries 01}
  (2017) 090},
\href{http://arxiv.org/abs/1610.03400}{{\ttfamily arXiv:1610.03400 [gr-qc]}}.
%%CITATION = ARXIV:1610.03400;%%.

\bibitem{Creminelli:2016zwa}
P.~Creminelli, D.~Pirtskhalava, L.~Santoni, and E.~Trincherini, ``{Stability of
  Geodesically Complete Cosmologies},''
  \href{http://dx.doi.org/10.1088/1475-7516/2016/11/047}{{\em JCAP} {\bfseries
  1611} no.~11, (2016) 047},
\href{http://arxiv.org/abs/1610.04207}{{\ttfamily arXiv:1610.04207 [hep-th]}}.
%%CITATION = ARXIV:1610.04207;%%.

\bibitem{Cai:2017tku}
Y.~Cai, H.-G. Li, T.~Qiu, and Y.-S. Piao, ``{The Effective Field Theory of
  nonsingular cosmology: II},''
  \href{http://dx.doi.org/10.1140/epjc/s10052-017-4938-y}{{\em Eur. Phys. J.}
  {\bfseries C77} no.~6, (2017) 369},
\href{http://arxiv.org/abs/1701.04330}{{\ttfamily arXiv:1701.04330 [gr-qc]}}.
%%CITATION = ARXIV:1701.04330;%%.

\bibitem{HP70}
S.~W. Hawking and R.~Penrose, ``{The Singularities of gravitational collapse
  and cosmology},''
{\em Proc Roy Soc Lond} {\bfseries A314} (1970) 529--548.
%%CITATION = PRSLA,A314,529;%%.

\bibitem{Creminelli:2006xe}
P.~Creminelli, M.~A. Luty, A.~Nicolis, and L.~Senatore, ``{Starting the
  Universe: Stable Violation of the Null Energy Condition and Non-standard
  Cosmologies},'' \href{http://dx.doi.org/10.1088/1126-6708/2006/12/080}{{\em
  JHEP} {\bfseries 12} (2006) 080},
\href{http://arxiv.org/abs/hep-th/0606090}{{\ttfamily arXiv:hep-th/0606090
  [hep-th]}}.
%%CITATION = HEP-TH/0606090;%%.

\bibitem{Creminelli:2007aq}
P.~Creminelli and L.~Senatore, ``{A Smooth bouncing cosmology with scale
  invariant spectrum},''
  \href{http://dx.doi.org/10.1088/1475-7516/2007/11/010}{{\em JCAP} {\bfseries
  0711} (2007) 010},
\href{http://arxiv.org/abs/hep-th/0702165}{{\ttfamily arXiv:hep-th/0702165
  [hep-th]}}.
%%CITATION = HEP-TH/0702165;%%.

\bibitem{BKO07}
E.~I. Buchbinder, J.~Khoury, and B.~A. Ovrut, ``{New Ekpyrotic cosmology},''
  {\em Phys.~Rev.~D} {\bfseries 76} (2007) 123503,
  \href{http://arxiv.org/abs/arXiv.org: hep-th/0702154}{{\ttfamily arXiv.org:
  hep-th/0702154}}.

\bibitem{Lehners:2011kr}
J.-L. Lehners, ``{Cosmic Bounces and Cyclic Universes},''
  \href{http://dx.doi.org/10.1088/0264-9381/28/20/204004}{{\em Class. Quant.
  Grav.} {\bfseries 28} (2011) 204004},
\href{http://arxiv.org/abs/1106.0172}{{\ttfamily arXiv:1106.0172 [hep-th]}}.
%%CITATION = ARXIV:1106.0172;%%.

\bibitem{Qiu:2011cy}
T.~Qiu, J.~Evslin, Y.-F. Cai, M.~Li, and X.~Zhang, ``{Bouncing Galileon
  Cosmologies},'' \href{http://dx.doi.org/10.1088/1475-7516/2011/10/036}{{\em
  JCAP} {\bfseries 1110} (2011) 036},
\href{http://arxiv.org/abs/1108.0593}{{\ttfamily arXiv:1108.0593 [hep-th]}}.
%%CITATION = ARXIV:1108.0593;%%.

\bibitem{Easson:2011zy}
D.~A. Easson, I.~Sawicki, and A.~Vikman, ``{G-Bounce},''
  \href{http://dx.doi.org/10.1088/1475-7516/2011/11/021}{{\em JCAP} {\bfseries
  1111} (2011) 021},
\href{http://arxiv.org/abs/1109.1047}{{\ttfamily arXiv:1109.1047 [hep-th]}}.
%%CITATION = ARXIV:1109.1047;%%.

\bibitem{Cai:2012va}
Y.-F. Cai, D.~A. Easson, and R.~Brandenberger, ``{Towards a Nonsingular
  Bouncing Cosmology},''
  \href{http://dx.doi.org/10.1088/1475-7516/2012/08/020}{{\em JCAP} {\bfseries
  1208} (2012) 020},
\href{http://arxiv.org/abs/1206.2382}{{\ttfamily arXiv:1206.2382 [hep-th]}}.
%%CITATION = ARXIV:1206.2382;%%.

\bibitem{Qiu:2013eoa}
T.~Qiu, X.~Gao, and E.~N. Saridakis, ``{Towards anisotropy-free and nonsingular
  bounce cosmology with scale-invariant perturbations},''
  \href{http://dx.doi.org/10.1103/PhysRevD.88.043525}{{\em Phys. Rev.}
  {\bfseries D88} no.~4, (2013) 043525},
\href{http://arxiv.org/abs/1303.2372}{{\ttfamily arXiv:1303.2372
  [astro-ph.CO]}}.
%%CITATION = ARXIV:1303.2372;%%.

\bibitem{Ijjas:2016tpn}
A.~Ijjas and P.~J. Steinhardt, ``{Classically stable nonsingular cosmological
  bounces},'' \href{http://dx.doi.org/10.1103/PhysRevLett.117.121304}{{\em
  Phys. Rev. Lett.} {\bfseries 117} no.~12, (2016) 121304},
\href{http://arxiv.org/abs/1606.08880}{{\ttfamily arXiv:1606.08880 [gr-qc]}}.
%%CITATION = ARXIV:1606.08880;%%.

\bibitem{Ijjas:2016vtq}
A.~Ijjas and P.~J. Steinhardt, ``{Fully stable cosmological solutions with a
  non-singular classical bounce},''
  \href{http://dx.doi.org/10.1016/j.physletb.2016.11.047}{{\em Phys. Lett.}
  {\bfseries B764} (2017) 289--294},
\href{http://arxiv.org/abs/1609.01253}{{\ttfamily arXiv:1609.01253 [gr-qc]}}.
%%CITATION = ARXIV:1609.01253;%%.

\bibitem{Qiu:2015nha}
T.~Qiu and Y.-T. Wang, ``{G-Bounce Inflation: Towards Nonsingular Inflation
  Cosmology with Galileon Field},''
  \href{http://dx.doi.org/10.1007/JHEP04(2015)130}{{\em JHEP} {\bfseries 04}
  (2015) 130},
\href{http://arxiv.org/abs/1501.03568}{{\ttfamily arXiv:1501.03568
  [astro-ph.CO]}}.
%%CITATION = ARXIV:1501.03568;%%.

\bibitem{Wan:2015hya}
Y.~Wan, T.~Qiu, F.~P. Huang, Y.-F. Cai, H.~Li, and X.~Zhang, ``{Bounce
  Inflation Cosmology with Standard Model Higgs Boson},''
  \href{http://dx.doi.org/10.1088/1475-7516/2015/12/019}{{\em JCAP} {\bfseries
  1512} no.~12, (2015) 019},
\href{http://arxiv.org/abs/1509.08772}{{\ttfamily arXiv:1509.08772 [gr-qc]}}.
%%CITATION = ARXIV:1509.08772;%%.

\bibitem{Koehn:2012te}
M.~Koehn, J.-L. Lehners, and B.~Ovrut, ``{Ghost condensate in $N=1$
  supergravity},'' \href{http://dx.doi.org/10.1103/PhysRevD.87.065022}{{\em
  Phys. Rev.} {\bfseries D87} no.~6, (2013) 065022},
\href{http://arxiv.org/abs/1212.2185}{{\ttfamily arXiv:1212.2185 [hep-th]}}.
%%CITATION = ARXIV:1212.2185;%%.

\bibitem{Koehn:2013upa}
M.~Koehn, J.-L. Lehners, and B.~A. Ovrut, ``{Cosmological super-bounce},''
  \href{http://dx.doi.org/10.1103/PhysRevD.90.025005}{{\em Phys. Rev.}
  {\bfseries D90} no.~2, (2014) 025005},
\href{http://arxiv.org/abs/1310.7577}{{\ttfamily arXiv:1310.7577 [hep-th]}}.
%%CITATION = ARXIV:1310.7577;%%.

\bibitem{Battarra:2014tga}
L.~Battarra, M.~Koehn, J.-L. Lehners, and B.~A. Ovrut, ``{Cosmological
  Perturbations Through a Non-Singular Ghost-Condensate/Galileon Bounce},''
  \href{http://dx.doi.org/10.1088/1475-7516/2014/07/007}{{\em JCAP} {\bfseries
  1407} (2014) 007},
\href{http://arxiv.org/abs/1404.5067}{{\ttfamily arXiv:1404.5067 [hep-th]}}.
%%CITATION = ARXIV:1404.5067;%%.

\bibitem{Koehn:2015vvy}
M.~Koehn, J.-L. Lehners, and B.~Ovrut, ``{Nonsingular bouncing cosmology:
  Consistency of the effective description},''
  \href{http://dx.doi.org/10.1103/PhysRevD.93.103501}{{\em Phys. Rev.}
  {\bfseries D93} no.~10, (2016) 103501},
\href{http://arxiv.org/abs/1512.03807}{{\ttfamily arXiv:1512.03807 [hep-th]}}.
%%CITATION = ARXIV:1512.03807;%%.

\bibitem{deRham:2017aoj}
C.~de~Rham and S.~Melville, ``{Unitary NEC violation in P(X) cosmologies},''
\href{http://arxiv.org/abs/1703.00025}{{\ttfamily arXiv:1703.00025 [hep-th]}}.
%%CITATION = ARXIV:1703.00025;%%.

\bibitem{ArmendarizPicon:2003qk}
C.~Armendariz-Picon and P.~B. Greene, ``{Spinors, inflation, and nonsingular
  cyclic cosmologies},'' \href{http://dx.doi.org/10.1023/A:1025783118888}{{\em
  Gen. Rel. Grav.} {\bfseries 35} (2003) 1637--1658},
\href{http://arxiv.org/abs/hep-th/0301129}{{\ttfamily arXiv:hep-th/0301129
  [hep-th]}}.
%%CITATION = HEP-TH/0301129;%%.

\bibitem{Hehl:1976kj}
F.~W. Hehl, P.~Von Der~Heyde, G.~D. Kerlick, and J.~M. Nester, ``{General
  Relativity with Spin and Torsion: Foundations and Prospects},''
\href{http://dx.doi.org/10.1103/RevModPhys.48.393}{{\em Rev. Mod. Phys.}
  {\bfseries 48} (1976) 393--416}.
%%CITATION = RMPHA,48,393;%%.

\bibitem{Ortin:2015hya}
T.~Ortin, {\em {Gravity and Strings}}.
\newblock Cambridge Monographs on Mathematical Physics. Cambridge University
  Press, 2015.
\newblock
\url{http://www.cambridge.org/mw/academic/subjects/physics/theoretical-physics-and-mathematical-physics/gravity-and-strings-2nd-edition}.
\newblock
%%CITATION = INSPIRE-1383727;%%.

\bibitem{Kibble:1961ba}
T.~W.~B. Kibble, ``{Lorentz invariance and the gravitational field},''
\href{http://dx.doi.org/10.1063/1.1703702}{{\em J. Math. Phys.} {\bfseries 2}
  (1961) 212--221}.
%%CITATION = JMAPA,2,212;%%.

\bibitem{Sciama:1964wt}
D.~W. Sciama, ``{The Physical structure of general relativity},''
  \href{http://dx.doi.org/10.1103/RevModPhys.36.1103}{{\em Rev. Mod. Phys.}
  {\bfseries 36} (1964) 463--469}.
[Erratum: Rev. Mod. Phys.36,1103(1964)].
%%CITATION = RMPHA,36,463;%%.

\bibitem{Stelle:1979va}
K.~S. Stelle and P.~C. West, ``{De Sitter Gauge Invariance and the Geometry of
  the Einstein-Cartan theory},''
\href{http://dx.doi.org/10.1088/0305-4470/12/8/003}{{\em J. Phys.} {\bfseries
  A12} (1979) L205--L210}.
%%CITATION = JPAGA,A12,L205;%%.

\bibitem{Poplawski:2012ab}
N.~J. Poplawski, ``{Big bounce from spin and torsion},''
  \href{http://dx.doi.org/10.1007/s10714-011-1323-2}{{\em Gen. Rel. Grav.}
  {\bfseries 44} (2012) 1007--1014},
\href{http://arxiv.org/abs/1105.6127}{{\ttfamily arXiv:1105.6127
  [astro-ph.CO]}}.
%%CITATION = ARXIV:1105.6127;%%.

\bibitem{Poplawski:2011jz}
N.~J. Poplawski, ``{Nonsingular, big-bounce cosmology from spinor-torsion
  coupling},'' \href{http://dx.doi.org/10.1103/PhysRevD.85.107502}{{\em Phys.
  Rev.} {\bfseries D85} (2012) 107502},
\href{http://arxiv.org/abs/1111.4595}{{\ttfamily arXiv:1111.4595 [gr-qc]}}.
%%CITATION = ARXIV:1111.4595;%%.

\bibitem{Magueijo:2012ug}
J.~Magueijo, T.~G. Zlosnik, and T.~W.~B. Kibble, ``{Cosmology with a spin},''
  \href{http://dx.doi.org/10.1103/PhysRevD.87.063504}{{\em Phys. Rev.}
  {\bfseries D87} no.~6, (2013) 063504},
\href{http://arxiv.org/abs/1212.0585}{{\ttfamily arXiv:1212.0585
  [astro-ph.CO]}}.
%%CITATION = ARXIV:1212.0585;%%.

\bibitem{Alexander:2014eva}
S.~Alexander, C.~Bambi, A.~Marciano, and L.~Modesto, ``{Fermi-bounce Cosmology
  and scale invariant power-spectrum},''
  \href{http://dx.doi.org/10.1103/PhysRevD.90.123510}{{\em Phys. Rev.}
  {\bfseries D90} no.~12, (2014) 123510},
\href{http://arxiv.org/abs/1402.5880}{{\ttfamily arXiv:1402.5880 [gr-qc]}}.
%%CITATION = ARXIV:1402.5880;%%.

\bibitem{Alexander:2014uaa}
S.~Alexander, Y.-F. Cai, and A.~Marciano, ``{Fermi-bounce cosmology and the
  fermion curvaton mechanism},''
  \href{http://dx.doi.org/10.1016/j.physletb.2015.04.026}{{\em Phys. Lett.}
  {\bfseries B745} (2015) 97--104},
\href{http://arxiv.org/abs/1406.1456}{{\ttfamily arXiv:1406.1456 [gr-qc]}}.
%%CITATION = ARXIV:1406.1456;%%.

\bibitem{Freidel:2005sn}
L.~Freidel, D.~Minic, and T.~Takeuchi, ``{Quantum gravity, torsion, parity
  violation and all that},''
  \href{http://dx.doi.org/10.1103/PhysRevD.72.104002}{{\em Phys. Rev.}
  {\bfseries D72} (2005) 104002},
\href{http://arxiv.org/abs/hep-th/0507253}{{\ttfamily arXiv:hep-th/0507253
  [hep-th]}}.
%%CITATION = HEP-TH/0507253;%%.

\bibitem{Rezende:2009sv}
D.~J. Rezende and A.~Perez, ``{4d Lorentzian Holst action with topological
  terms},'' \href{http://dx.doi.org/10.1103/PhysRevD.79.064026}{{\em Phys.
  Rev.} {\bfseries D79} (2009) 064026},
\href{http://arxiv.org/abs/0902.3416}{{\ttfamily arXiv:0902.3416 [gr-qc]}}.
%%CITATION = ARXIV:0902.3416;%%.

\bibitem{Randono:2005up}
A.~Randono, ``{A Note on parity violation and the Immirzi parameter},''
\href{http://arxiv.org/abs/hep-th/0510001}{{\ttfamily arXiv:hep-th/0510001
  [hep-th]}}.
%%CITATION = HEP-TH/0510001;%%.

\bibitem{Khriplovich:2005jh}
I.~B. Khriplovich and A.~A. Pomeransky, ``{Remark on Immirzi parameter,
  torsion, and discrete symmetries},''
  \href{http://dx.doi.org/10.1103/PhysRevD.73.107502}{{\em Phys. Rev.}
  {\bfseries D73} (2006) 107502},
\href{http://arxiv.org/abs/hep-th/0508136}{{\ttfamily arXiv:hep-th/0508136
  [hep-th]}}.
%%CITATION = HEP-TH/0508136;%%.

\bibitem{Ellis:2011mz}
J.~Ellis and N.~E. Mavromatos, ``{On the Role of Space-Time Foam in Breaking
  Supersymmetry via the Barbero-Immirzi Parameter},''
  \href{http://dx.doi.org/10.1103/PhysRevD.84.085016}{{\em Phys. Rev.}
  {\bfseries D84} (2011) 085016},
\href{http://arxiv.org/abs/1108.0877}{{\ttfamily arXiv:1108.0877 [hep-th]}}.
%%CITATION = ARXIV:1108.0877;%%.

\bibitem{Dolan:2009ni}
B.~P. Dolan, ``{Chiral fermions and torsion in the early Universe},''
  \href{http://dx.doi.org/10.1088/0264-9381/27/9/095010,
  10.1088/0264-9381/27/24/249801}{{\em Class. Quant. Grav.} {\bfseries 27}
  (2010) 095010}, \href{http://arxiv.org/abs/0911.1636}{{\ttfamily
  arXiv:0911.1636 [gr-qc]}}.
[Erratum: Class. Quant. Grav.27,249801(2010)].
%%CITATION = ARXIV:0911.1636;%%.

\bibitem{Isham:1974ci}
C.~J. Isham and J.~E. Nelson, ``{Quantization of a Coupled Fermi Field and
  Robertson-Walker Metric},''
\href{http://dx.doi.org/10.1103/PhysRevD.10.3226}{{\em Phys. Rev.} {\bfseries
  D10} (1974) 3226}.
%%CITATION = PHRVA,D10,3226;%%.

\bibitem{Erickson:2003zm}
J.~K. Erickson, D.~H. Wesley, P.~J. Steinhardt, and N.~Turok, ``{Kasner and
  mixmaster behavior in universes with equation of state w >= 1},''
  \href{http://dx.doi.org/10.1103/PhysRevD.69.063514}{{\em Phys. Rev.}
  {\bfseries D69} (2004) 063514},
\href{http://arxiv.org/abs/hep-th/0312009}{{\ttfamily arXiv:hep-th/0312009
  [hep-th]}}.
%%CITATION = HEP-TH/0312009;%%.

\bibitem{Lehners:2008vx}
J.-L. Lehners, ``{Ekpyrotic and Cyclic Cosmology},''
  \href{http://dx.doi.org/10.1016/j.physrep.2008.06.001}{{\em Phys. Rept.}
  {\bfseries 465} (2008) 223--263},
\href{http://arxiv.org/abs/0806.1245}{{\ttfamily arXiv:0806.1245 [astro-ph]}}.
%%CITATION = ARXIV:0806.1245;%%.

\bibitem{Lehners:2013cka}
J.-L. Lehners and P.~J. Steinhardt, ``{Planck 2013 results support the cyclic
  universe},'' \href{http://dx.doi.org/10.1103/PhysRevD.87.123533}{{\em Phys.
  Rev.} {\bfseries D87} no.~12, (2013) 123533},
\href{http://arxiv.org/abs/1304.3122}{{\ttfamily arXiv:1304.3122
  [astro-ph.CO]}}.
%%CITATION = ARXIV:1304.3122;%%.

\bibitem{Ijjas:2015hcc}
A.~Ijjas and P.~J. Steinhardt, ``{Implications of Planck2015 for inflationary,
  ekpyrotic and anamorphic bouncing cosmologies},''
  \href{http://dx.doi.org/10.1088/0264-9381/33/4/044001}{{\em Class. Quant.
  Grav.} {\bfseries 33} no.~4, (2016) 044001},
\href{http://arxiv.org/abs/1512.09010}{{\ttfamily arXiv:1512.09010
  [astro-ph.CO]}}.
%%CITATION = ARXIV:1512.09010;%%.

\bibitem{Baumann:2009ds}
D.~Baumann,
  \href{http://dx.doi.org/10.1142/9789814327183_0010}{``{Inflation},''} in {\em
  {Physics of the large and the small, TASI 09, proceedings of the Theoretical
  Advanced Study Institute in Elementary Particle Physics, Boulder, Colorado,
  USA, 1-26 June 2009}}, pp.~523--686.
\newblock 2011.
\newblock \href{http://arxiv.org/abs/0907.5424}{{\ttfamily arXiv:0907.5424
  [hep-th]}}.
\newblock
\url{https://inspirehep.net/record/827549/files/arXiv:0907.5424.pdf}.
\newblock
%%CITATION = ARXIV:0907.5424;%%.

\bibitem{Damour:2011yk}
T.~Damour and P.~Spindel, ``{Quantum Einstein-Dirac Bianchi Universes},''
  \href{http://dx.doi.org/10.1103/PhysRevD.83.123520}{{\em Phys. Rev.}
  {\bfseries D83} (2011) 123520},
\href{http://arxiv.org/abs/1103.2927}{{\ttfamily arXiv:1103.2927 [gr-qc]}}.
%%CITATION = ARXIV:1103.2927;%%.

\bibitem{Boyle:2003km}
L.~A. Boyle, P.~J. Steinhardt, and N.~Turok, ``{The Cosmic gravitational wave
  background in a cyclic universe},''
  \href{http://dx.doi.org/10.1103/PhysRevD.69.127302}{{\em Phys. Rev.}
  {\bfseries D69} (2004) 127302},
\href{http://arxiv.org/abs/hep-th/0307170}{{\ttfamily arXiv:hep-th/0307170
  [hep-th]}}.
%%CITATION = HEP-TH/0307170;%%.

\bibitem{Stewart:1990fm}
J.~M. Stewart, ``{Perturbations of Friedmann-Robertson-Walker cosmological
  models},''
\href{http://dx.doi.org/10.1088/0264-9381/7/7/013}{{\em Class. Quant. Grav.}
  {\bfseries 7} (1990) 1169--1180}.
%%CITATION = CQGRD,7,1169;%%.

\end{thebibliography}\endgroup

\end{document}